\documentclass[oneeqnum,onefignum,onetabnum,onethmnum]{siamltex}
\usepackage{graphics}

\newcommand{\bc}{\begin{center}}
\newcommand{\ec}{\end{center}}
\newcommand{\beq}{\begin{equation}}
\newcommand{\eeq}{\end{equation}}
\newcommand{\bea}{\begin{eqnarray}}
\newcommand{\eea}{\end{eqnarray}}

\def\e23{\epsilon^{2/3}}
\def\cb{\bar{c}}
\def\rhob{{\bar{\rho}}}
\def\Eb{{\bar{E}}}
\def\phib{{\bar{\phi}}}

\def\ch{\hat{c}}
\def\rhoh{{\hat{\rho}}}

\def\phih{{\hat{\phi}}}

\def\cc{\check{c}}
\def\rhoc{{\check{\rho}}}

\def\phic{{\check{\phi}}}

\def\i{j}
\def\ir{\i_r}
\def\I{I}
\def\ID{\I_d}

\title{Current-voltage relations for electrochemical thin films}

\author{
Martin Z. Bazant\footnotemark[2],
Kevin T. Chu\footnotemark[2], 
\and B. J. Bayly\footnotemark[3]}

\date{ June 12, 2004 }    

\begin{document}
\maketitle

\renewcommand{\thefootnote}{\fnsymbol{footnote}}
\footnotetext[2]{Department of Mathematics, Massachusetts Institute of
Technology, Cambridge, MA 02139}
\footnotetext[3]{Department of Mathematics, University of Arizona,
Tucson, AZ 85271}
\renewcommand{\thefootnote}{\arabic{footnote}}

\begin{abstract}
The dc response of an electrochemical thin film, such as the separator
in a micro-battery, is analyzed by solving the Poisson-Nernst-Planck
equations, subject to boundary conditions appropriate for an
electrolytic/galvanic cell.  The model system consists of a binary
electrolyte between parallel-plate electrodes, each possessing a
compact Stern layer, which mediates Faradaic reactions with nonlinear
Butler-Volmer kinetics. Analytical results are obtained by matched
asymptotic expansions in the limit of thin double layers and compared
with full numerical solutions.  The analysis shows that (i) decreasing
the system size relative to the Debye screening length decreases the
voltage of the cell and allows currents higher than the classical
diffusion-limited current; (ii) finite reaction rates lead to the
important possibility of a reaction-limited current; (iii) the
Stern-layer capacitance is critical for allowing the cell to achieve
currents above the reaction-limited current; and (iv) all
polarographic (current-voltage) curves tend to the same limit as
reaction kinetics become fast.  Dimensional analysis, however, shows
that ``fast'' reactions tend to become ``slow'' with decreasing system
size, so the nonlinear effects of surface polarization
may dominate the dc response of thin films.
\end{abstract}



\pagestyle{myheadings}
\thispagestyle{plain}
\markboth{MARTIN Z. BAZANT, KEVIN T. CHU, AND B.~J. BAYLY}
{ELECTROCHEMICAL CURRENT-VOLTAGE RELATIONS}

\section*{Introduction}
Micro-electrochemical systems pose interesting problems for applied
mathematics because traditional ``macroscopic'' approximations of
electroneutrality and thermal equilibrium~\cite{newman_book}, which
make the classical transport equations more
tractable~\cite{rubinstein_book}, break down at small scales,
approaching the Debye screening length.  Of course, the relative
importance of surface phenomena also increases with miniaturization.
Micro-electrochemical systems of current interest include ion channels
in biological membranes~\cite{barcilon1992,park1997,barcilon1997} and
thin-film
batteries~\cite{dudney1995,wang1996,neudecker2000,takami2002,shi2003},
which could revolutionize the design of modern electronics with
distributed on-chip power sources.  In the latter context, the
internal resistance of the battery is related to the nonlinear
current-voltage characteristics of the separator, consisting of a
thin-film electrolyte (solid, liquid, or gel) sandwiched between flat
electrodes and interfacial layers where Faradaic electron-transfer
reactions occur~\cite{brett_book}. Under such conditions, the internal
resistance is unlikely to be simply constant, as is usually 
assumed.

Motivated by the application to thin-film batteries, here we revisit
the classical problem of steady conduction between parallel, flat
electrodes, studied by Nernst~\cite{nernst1904} and
Brunner~\cite{brunner1904,brunner1907} a century ago.  As in
subsequent studies of liquid~\cite{jaffe1952,jaffe1953} and
solid~\cite{itskovich1977,kornyshev1981} electrolytes, we do not make
Nernst's assumption of bulk electroneutrality and work instead with
the Poisson-Nernst-Planck (PNP) equations, allowing for diffuse charge
in solution~\cite{newman_book,rubinstein_book}. What distinguishes our
analysis from previous work on current-voltage relations (or
``polarographic curves'') is the use of more realistic nonlinear
boundary conditions describing (i) Butler-Volmer reaction kinetics and
(ii) the surface capacitance of the compact Stern layer, as in the
recent paper of Bonnefont {\it et. al}~\cite{bonnefont2001}. Such
boundary conditions, although complicating mathematical analysis,
generally cannot be ignored in micro-electrochemical cells, where
interfaces play a crucial role. Diffuse-charge dynamics, which can be
important for high-power applications, also complicates
analysis~\cite{bazant2004}, but in most cases it is reasonable to
assume that electrochemical thin films are in steady state, due to the
short distances for electro-diffusion.

We stress, however, that steady state does not imply thermal
equilibrium when the system sustains a current, driven by an applied
voltage. In this paper, we focus on applied voltages small enough to
justify the standard boundary-layer analysis of the PNP equations,
which yields charge densities in thermal equilibrium at leading order
in the limit of thin double layers~\cite{newman1965}. In a companion
paper~\cite{chu2004}, we extend the analysis to larger voltages,
at~\cite{smyrl1967} and above~\cite{rubinstein1979} the Nernst's
diffusion-limited current, and show how more realistic boundary
conditions affect diffuse charge, far from thermal equilibrium. In
both cases, we obtain novel formulae for polarographic curves by
asymptotic analysis in the limit of thin double layers, which we
compare with numerical solutions, and we focus on a 
variety of dimensionless physical parameters: the reaction-rate
constants (scaled to the typical diffusive flux) and the ratios of the
Stern length to the Debye length to the electrode separation.

\section{Mathematical Model} Let us consider  uniform 
conduction through a dilute, binary electrolyte between parallel plate
electrodes separated by a distance, $L$.  Our goal is to determine the
steady-state response of the electrochemical cell to either an applied
voltage, $V$, or an applied current, $\I$.  Specifically, we seek the
electric potential $\Phi(X)$ and the concentrations $C_+(X)$ and
$C_-(X)$ of cations and anions in the region $0 \leq X \leq L$.  The
Faradaic current is driven by redox reactions occurring at the
electrodes, and we neglect any other chemical reactions, such as
dissociation/recombination in the bulk solution or hydrogen
production.  Since we do not assume electroneutrality, the region of
integration extends to the point where the continuum approximation
breaks down near each electrode, roughly a few molecules away. In
other words, our integration region includes the ``diffuse part'' but
not the ``compact part'' of the double layer 
\cite{newman_book,brett_book,delahay_book,bard_book}.

\subsection{ Transport Equations.} In the context of dilute solution
theory \cite{newman_book,rubinstein_book}, the governing 
equations for this situation are the steady PNP equations:
\bea
\frac{d}{dX}\left( D_+ \frac{dC_+}{dX} + 
	\mu_+z_+F C_+ \frac{d\Phi}{dX}\right)  = 
	0   \label{eq:cationeq}  \\
\frac{d}{dX}\left( D_- \frac{dC_-}{dX} + \mu_-z_-F
C_- \frac{d\Phi}{dX}\right)  =  0   \label{eq:anioneq} \\
	-\frac{d}{dX}\left(\epsilon_s \frac{d\Phi}{dX}\right) 
	= (z_+C_+ - z_-C_-)F   
	\label{eq:poissoneq} . 
\eea
where $F$ is Faraday's constant (a mole of charge), $z_\pm$,
$\mu_\pm$, and $D_\pm$ are the charge numbers, mobilities, and
diffusivities of each ionic species, respectively, and $\epsilon_s$ is
the permittivity of the solvent, all taken to be constant in the limit
of infinite dilution.  The first two equations set divergences of the
ionic fluxes to zero in order to maintain the steady state, and the
third is Poisson's equation relating the electric potential to the
charge density. In each flux expression, the first term represents
diffusion and the second electromigration. The Einstein relation,
$\mu_\pm = D_\pm/RT$, relates mobilities and diffusion coefficients
via the absolute temperature, $T$, and the universal gas constant,
$R$.  

Due to the potentially large electric fields in thin films, as in
interfacial double layers, these classical approximations
could break down~\cite{newman_book,delahay_book,bard_book}. For
example, the polarization of solvent molecules in large electric
fields can lower the solvent dielectric permittivity by an order of
magnitude, while also affecting the diffusivity and mobility. The
solution can also become locally so concentrated or depleted of
certain ions as to make finite sizes and/or interactions (and thus
ionic activities) important.  In spite of these concerns, however, it
is reasonable to analyze the PNP equations before considering more
complicated transport models, especially because our focus is on the
effect of boundary conditions.

\subsection{Electrode Boundary Conditions} Although PNP equations
constitute a well-understood and widely accepted approximation,
appropriate boundary conditions for them are not so clear, and drastic
approximations, such as constant concentration, potential or surface
charge (or zeta potential), are usually made, largely out of
mathematical convenience.  On the other hand, in the context of
electric circuit models for electrochemical
cells~\cite{bazant2004,macdonald1990,geddes1997}, much effort has been
made to describe the nonlinear response of the electrode-electrolyte
interface, while describing the bulk solution as a simple circuit
element, such as resistor. Here, we formulate general boundary
conditions based on classical models of the double
layer~\cite{brett_book,delahay_book,bard_book}, with a unified
description of ion transport by the PNP equations.

Our first pair of boundary conditions sets the normal anion flux 
to zero at each electrode,
\bea
D_- \frac{dC_-}{dX}(0) + \mu_- z_-F
C_-(0) \frac{d\Phi}{dX}(0) &=& 0 \label{eq:anbc} \\
D_- \frac{dC_-}{dX}(L) + \mu_- z_-F
C_-(L) \frac{d\Phi}{dX}(L) &=& 0 ,
\eea
on the assumption that anions do not specifically adsorb onto the
surfaces, which holds for many anions at typical metal surfaces
(e.g. SO$_4^{-2}$, OH$^-$, F$^-$).  The second pair relates the normal
cation flux to the net deposition (or dissolution) flux, or
reaction-rate density, $R(C_+,\Delta\Phi_S)$, which in the dilute
limit is assumed to depend only on cation concentration and potential
drop, $\Delta\Phi_S$, across the compact part of the double layer,
originally proposed by Stern~\cite{stern1924}.  Following the
convention in electrochemistry, we take $\Delta\Phi_S$ to be the
potential of the electrode surface measured relative to the
solution.  The reference potential is
chosen so that the cathode, located at $X=0$, is at zero potential and
the anode, located at $X=1$, is at the applied cell voltage
$V$. Therefore, we have the following two boundary conditions,
\bea
D_+ \frac{dC_+}{dX}(0) + \mu_+z_+F
C_+(0) \frac{d\Phi}{dX}(0) &=& R\left(C_+(0),\Phi(0)\right) 
\label{eq:catbcc} 
\\
-D_+ \frac{dC_+}{dX}(L) - \mu_+z_+F
C_+(L) \frac{d\Phi}{dX}(L) &=& R\left(C_+(L),\Phi(L)-V)\right) .
\label{eq:catbca}
\eea
For electrodes, it is typical to assume a balance of forward
(deposition) and backward (dissolution) reaction rates biased by the
Stern voltage with an Arrhenius temperature dependence,
\beq
R(C_+,\Delta\Phi_S) = K_c C_+ \exp\left(\frac{-\alpha_c z F
\Delta\Phi_S}{RT}\right) - K_a C_M \exp\left(\frac{\alpha_a
zF \Delta\Phi_S}{RT}\right) ,
\label{eq:bv}
\eeq
where $C_M$ is the (constant) density of electrode metal and $K_c$ and
$K_a$ are rate constants for the cathodic and anodic reactions 
\cite{newman_book}.  (Expressing the reaction rate in terms of the
surface overpotential,  $\eta_S = \Delta\Phi_S -
\Delta\Phi_S^{eq}$, where $\Delta\Phi_S^{eq}$ is the Stern-layer
voltage in the absence of current, $R=0$, yields the more common form
of the Butler-Volmer equation~\cite{newman_book,bard_book}.)
The Stern layer voltage contributes $-z F \Delta\Phi_S$ to the
activation energy barriers multiplied by transfer coefficients
$\alpha_c$ and $\alpha_a$ for the cathodic and anodic reactions,
respectively, where $\alpha_c \approx \alpha_a \approx
\frac{1}{2}$, for single electron transfer
reactions~\cite{newman_book,bard_book,chidsey_article}.  

Following Frumkin \cite{frumkin}, we apply Eq.~(\ref{eq:bv}) just
outside the Stern layer, in contrast to ``macroscopic" models which
postulate the Butler-Volmer equation as a purely empirical description
of reactions between the electrode surface and the electrically
neutral bulk solution.  Physically, the Frumkin approach makes more
sense since the activation energy barrier described by the
Butler-Volmer equation actually exists at the atomic scale in the
Stern layer, not across the entire ``interface'' including diffuse
charge in solution.
We are not aware of any prior analysis with the full, nonlinear
Butler-Volmer equation as a boundary condition on the PNP equations
other than that of Bonnefont {\it et
al.}~\cite{bonnefont2001}.  Earlier analyses by Jaff\'e {\it
et.~al.}\cite{jaffe1952,jaffe1953} and Itskovich {\it
et.~al.}\cite{itskovich1977} also include electrode reactions, but
only for small perturbations around equilibrium.

The final pair of boundary conditions determines the
electric potential by specifying the voltage drop across the
Stern layer in terms of the local electric field and
concentrations,
\bea
0 - \Phi(0) & = &
\Delta\Phi_S \left(\frac{d\Phi}{dX}(0),C_+(0),C_-(0)\right) 
	\label{eq:potc}\\ 
V - \Phi(L) & = &
\Delta\Phi_S \left(-\frac{d\Phi}{dX}(L),C_+(L),C_-(L)\right) .
	\label{eq:pota} 
\eea 
In macroscopic electrochemistry, these boundary conditions are usually
replaced by the assuption of electroneutrality (which eliminates the
need to solve Poisson's equation) or by simple Dirichlet boundary
conditions on the potential~\cite{newman_book}. In colloidal science,
they are likewise replaced by simple boundary conditions of constant
surface charge (or zeta
potential)~\cite{hunter_book,lyklema_book}. Here, we incorporate more
realistic properties of the interface as follows: Neglecting the
specific adsorption of anions, the Stern layer acts as a nonlinear
capacitor in series with the diffuse layer.  Grahame's celebrated
electrocapillary measurements~\cite{grahame1947,grahame1954} suggest
that $(i)$ the Stern layer capacitance, $C_S$, is roughly independent
of concentration, depending mainly on the (variable) total charge, $\sigma$,
\beq
\frac{d (\Delta\Phi_S)}{d\sigma} = \frac{1}{C_S(\sigma)} ,
\label{eq:cap}
\eeq
and $(ii)$ dilute solution theory accurately describes the
capacitance, $C_D(\sigma,C_+)$, of the diffuse layer, at least when
the charge and current are small enough to be well-described by
Poisson-Boltzmann theory (as derived below). Using Gauss' law, the
surface charge density can be expressed in terms of the normal
electric field, $\sigma = -\epsilon_S d\Phi/dX$, where $\epsilon_S$ is
an effective permittivity of the compact layer. Therefore, integrating
Eq.~(\ref{eq:cap}), Grahame's model corresponds to the assumption,
\beq 
\Delta\Phi_S = \int_0^{-\epsilon_S d\Phi/dx}
\frac{d\sigma}{C_S(\sigma)}, \label{eq:grahame}
\eeq
which determines how the voltage across the compact layer (relative to
the point of zero charge for which $\Delta\Phi_S = 0$) varies as the
two capacitors become charged.  The function, $C_S(\sigma)$, should be
fit to experimental or theoretical electrocapillary curves
at large concentrations (since $1/C_{total} = 1/C_D + 1/C_S \approx
1/C_S$ in that case).

The simplest model that captures this interplay between the compact
and diffuse layers is the Stern model~\cite{brett_book,bard_book},
which assumes the capacitance of the compact layer, $C_S$, to be
constant~\cite{stern1924}.  While more complicated models for the
compact layer have been
proposed~\cite{macdonald1954_b,macdonald1958,delahay_book}, the Stern
model suffices for our purposes, because it allows us to describe
surface capacitance easily in the context of our model of Faradaic
reactions.  Following Itskovich {\it et al.}~\cite{itskovich1977} and
Bonnefont {\it et al.}~\cite{bonnefont2001}, let us introduce an
effective width, $\lambda_S$, for the compact layer, $\lambda_S =
\epsilon_S/C_S$, so that Eq.~(\ref{eq:grahame}) reduces to a linear
extrapolation of the potential across the compact layer,
$-\Delta\Phi_S = \lambda_S d\Phi/dx$.  Substituting this expression
into Eqs.~(\ref{eq:potc}) and (\ref{eq:pota}) yields two Robin
boundary conditions,
\bea
\Phi(0) - \lambda_S \frac{d\Phi}{dX}(0) & = & 0 \\
\Phi(L) + \lambda_S \frac{d\Phi}{dX}(L) & = & V,
\eea
completing a set of six boundary conditions for our three second-order
differential equations.  Physically, the Stern layer, as an effective
solvation shell for the electrode, is only a few molecules wide, so it
is best to think of $\lambda_S$ as simply a measure of the capacitance
of the Stern layer.  More generally, the same boundary condition could
also describe a thin dielectric layer on the
electrode~\cite{ajdari2000,iceo2004a,iceo2004b}, e.g. arising from
surface contamination or a passivating monolayer.

Note that since the anion flux is zero, the current passing through the
cell is proportional to the cation flux (everywhere in the cell, since
it is constant),
\beq
\I = z_+ F A \left( D_+ \frac{dC_+}{dX} +
        \mu_+z_+F C_+ \frac{d\Phi}{dX}\right), 
\label{eq:current}
\eeq 
where $A$ is the electrode area and a current flow towards the cathode 
($x=0$) is taken to be positive.
Under potentiostatic conditions,
the cell voltage $V$ is given, and the steady-state polarization
curve $\I(V)$ is determined by solving the equations. Conversely,
under galvanostatic conditions, $\I$ is fixed, and we solve for
$V(\I)$.

\subsection{An Integral Constraint} As formulated above, the boundary
value problem is not well-posed.  Since the anion flux is constant
throughout the cell according to Eq.~(\ref{eq:anioneq}), the two
anion flux boundary conditions are degenerate, leaving one
constant of integration undetermined. 
This is not surprising as we have omitted one crucial physical
parameter, the total number of anions. 
More precisely, because anions do not react, 
we must specify their total number, which 
remains constant as the steady state is reached. This corresponds 
to the constraint,
\beq
\frac{1}{L} \int_0^L C_-(X) dX = C_{ref} ,
\eeq
where $C_{ref}$ is the initial concentration of anions.  

Note that the total number of cations (and hence the total charge) is
not known {\it a priori} because the removal of cations at the cathode
and the injection of cations at the anode may significantly alter
their total number.  This may seem counter-intuitive since we are
accustomed to assuming that we know the total cation concentration at
all times based on the original molarity of the solution, but this
``macroscopic" thinking does not apply when the physics at the
microscopic level are explicitly being studied ({\it e.g.} diffuse
charge layers or micro-electrochemical systems).  Mathematically, the
reaction boundary conditions at the electrodes Eqs.~(\ref{eq:catbcc})
and (\ref{eq:catbca}) are sufficient to determine the total cation
concentration (and total charge), as long as the total anion
concentration is specified.

\subsection{Dimensionless Formulation} To facilitate our analysis, we
formulate the problem in dimensionless form.  
For simplicity we also assume that
the electrolyte is symmetric, $z_+ = -z_- \equiv z$, which does not
qualitatively affect any of our conclusions as long as $z_+/|z_-|$ is not
too different from $1$ (which holds for most simple, aqueous electrolytes). 
Scaling the basic variables as follows,
\beq 
x \equiv X/L,\ \ \ c_\pm(x) \equiv
C_\pm(xL)/C_{ref}, \ \ \ \phi(x) \equiv \frac{zF\Phi(xL)}{RT},
\eeq
Eqs.~(\ref{eq:cationeq})--(\ref{eq:poissoneq}) become
\bea
\frac{d^2c_+}{dx^2} + \frac{d}{dx}\left( c_+ \frac{d\phi}{dx}
\right) = 0 \\
\frac{d^2c_-}{dx^2} - \frac{d}{dx}\left( c_- \frac{d\phi}{dx}
\right) = 0 \\
- \epsilon^2 \frac{d^2\phi}{dx^2} = \frac{1}{2}(c_+ - c_-) 
\eea
where $\epsilon \equiv \lambda_D/L$ is the ratio of the Debye
screening length 
$ \lambda_D \equiv \sqrt{\frac{\epsilon_sRT}{2z^2F^2C_{ref}} }$
to the distance between electrodes. The Debye length is typically
on the order of nanometers, so $\epsilon$ is always extremely
small for macroscopic electrochemical cells. This situation
changes, however, as $L$ or $C_{ref}$ is decreased, and in the
case of nano-electrochemical systems $\epsilon$ could be as large as $10$.

The two flux equations are easily integrated, using
Eq.~(\ref{eq:anbc}) to evaluate one constant and leaving the other
constant expressed in terms of the current via
Eq.~(\ref{eq:current}),
\bea
\frac{dc_+}{dx} + c_+ \frac{d\phi}{dx} & = & 4 \i 
  \label{eq:c+eq}\\
\frac{dc_-}{dx} - c_-\frac{d\phi}{dx} & = & 0 \label{eq:c-eq},
\eea
where we have defined a dimensionless current density, $\i \equiv \I/\ID$, 
scaled to the Nernst's diffusion-limited current density 
(see Section \ref{sec:bulk_soln_below_lim_cur}),
$\ID \equiv 4zFD_+C_{ref}A/L$.
Since diffuse charge is of primary interest here, it is convenient to
introduce 
\bea
c = \frac{1}{2}(c_+ + c_-) \ \ \ \textrm{and} \ \ \ 
\rho = \frac{1}{2}(c_+ - c_-) \label{eq:c_and_rho_def},
\eea
the average concentration of ions and (half) the charge density,
respectively, which leaves us with a coupled set of one second-order
and two first-order differential equations,
\bea
\frac{dc}{dx} + \rho \frac{d\phi}{dx} = 2 \i \label{eq:cphieq} \\
\frac{d\rho}{dx} + c \frac{d\phi}{dx} = 2 \i \label{eq:rhophieq} \\
- \epsilon^2 \frac{d^2\phi}{dx^2} = \rho. \label{eq:phieq} 
\eea
Nondimensionalizing the boundary conditions (remembering that
there is an extra condition needed to determine the current-voltage 
relation, $\i(v)$ or $v(\i)$), we obtain: 
\bea
\phi(0) - \delta \epsilon \frac{d\phi}{dx}(0) & = & 0 ,
\label{eq:potbc0} \\ 
\phi(1) + \delta \epsilon \frac{d\phi}{dx}(1) & = & v
\label{eq:potbc1} , \\
k_c[c(0) + \rho(0)]e^{\alpha_c\phi(0)} - \ir e^{-\alpha_a\phi(0)} & = & \i 
 \label{eq:potbvbc0} , \\
-k_c[c(1) + \rho(1)]e^{\alpha_c(\phi(1)-v)} + \ir
e^{-\alpha_a(\phi(1)-v)} & = & \i \label{eq:potbvbc1} ,  \\
\int_0^1 [c(x) - \rho(x)]dx & = & 1  ,\label{eq:cint}
\eea
where 
\begin{equation}
k_c \equiv \frac{K_cL}{4D_+}, \ \ \ \ir \equiv \frac{K_a L C_M}{4D_+C_{ref}}, \ \ \ v
\equiv \frac{zFV}{RT}, \ \ \mbox{ and} \  \delta \equiv
\frac{\lambda_S}{\lambda_D} .
\end{equation}
Keep in mind
that the dimensionless rate constants decrease with system size, so
that ``fast reactions'' ($k_c, \ir \gg 1$) may become ``slow
reactions'' ($k_c, \ir = O(1)$) as $L$ is reduced to the micron or
submicron scale.

It is important to note that we have scaled the effective Stern layer
width, $\lambda_S$, with the Debye screening length, $\lambda_D$,
rather than the electrode separation $L$, thus introducing the factor
$\epsilon = \lambda_D/L$ in Eqs.~(\ref{eq:potbc0}) and
(\ref{eq:potbc1}).  This choice is important for our asymptotic
analysis of the limit $\epsilon
\rightarrow 0$ at fixed $\delta$, which is intended to describe
situations in which $L$ is much larger than {\it both} $\lambda_S$ and
$\lambda_D$. Without it, our analysis would assume that as $\epsilon
\rightarrow 0$ the Stern layer becomes infinitely wide compared to the
diffuse layer, even though it is mainly the macroscopic electrode
separation which varies.  The limit of very small Stern layer
capacitance, which amounts to the Helmoltz model of the double layer
\cite{helmholtz1879}, is best studied by letting $\delta \rightarrow
\infty$ {\it after} $\epsilon
\rightarrow 0$. In contrast, because $\epsilon$ and $\delta$ would
both be small, the limit of very large Stern layer capacitance can be
studied by simply letting $\delta = 0$, yielding the Dirichlet boundary
conditions,
\beq
\phi(0) = 0 \ \ \ , \ \ \ 
\phi(1) = 1 , \label{eq:gcbc}
\eeq
of the Gouy-Chapman model of the double layer
\cite{gouy1910,chapman1913}. In our work, we shall consider both
limits, starting with the assumption that $\delta = O(1)$, which
corresponds to the Gouy-Chapman-Stern model of the double
layer~\cite{brett_book,bard_book}.

For one-dimensional problems, galvanostatic conditions are more
mathematically convenient than potentiostatic conditions. In the
former case, $\i$ is given, and $v(\i)$ is easily obtained from the
Stern boundary condition at the anode Eq.~(\ref{eq:potbc1}).  In the
latter case, however, $v$ is specified and $\i(v)$ must be determined
self-consistently to satisfy Eq.~(\ref{eq:potbc1}).  Therefore, we
shall assume that the current $\i$ is specified and solve
Eqs.~(\ref{eq:cphieq})--(\ref{eq:phieq}) subject to the boundary
conditions Eqs.~(\ref{eq:potbvbc0}) and (\ref{eq:potbvbc1}) and the
integral constraint Eq.~(\ref{eq:cint}).

\section{Boundary-layer Analysis}
In this section, we briefly review the classical asymptotic analysis
of the PNP equations, pioneered independently by
Chernenko~\cite{chernenko1963}, Newman~\cite{newman1965}, and
MacGillivray~\cite{macgillivray1968}, which involves boundary layers
of width, $\epsilon$ (corresponding to diffuse-charge layers of
dimensional width, $\lambda_D$). As discussed below, the classical
asymptotics breaks down at large currents approaching diffusion
limitation.  Unlike most previous authors, who assume either a fixed
potential~\cite{gouy1910,chapman1913} or fixed interfacial charge
\cite{newman1965} at an isolated electrode or fixed concentrations at cell 
boundaries with ion-permeable
membranes~\cite{park1997,barcilon1997,rubinstein1979}, we solve for
the response of a complete, two-electrode galvanic cell with boundary
conditions for Faradaic reactions and Stern-layer capacitance.

\begin{figure}
\bc
\scalebox{0.35}{\includegraphics{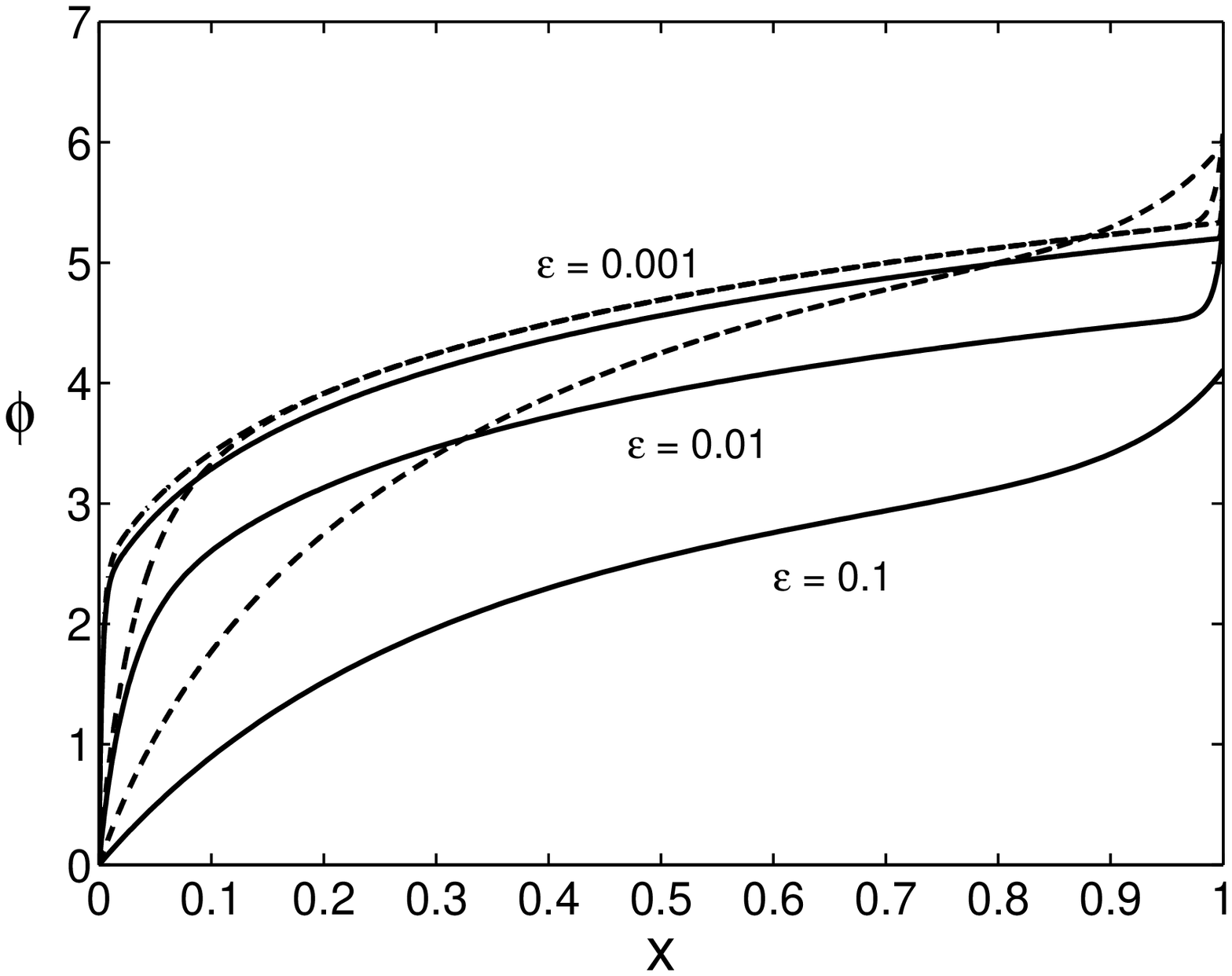}}
\ \nolinebreak 
\scalebox{0.35}{\includegraphics{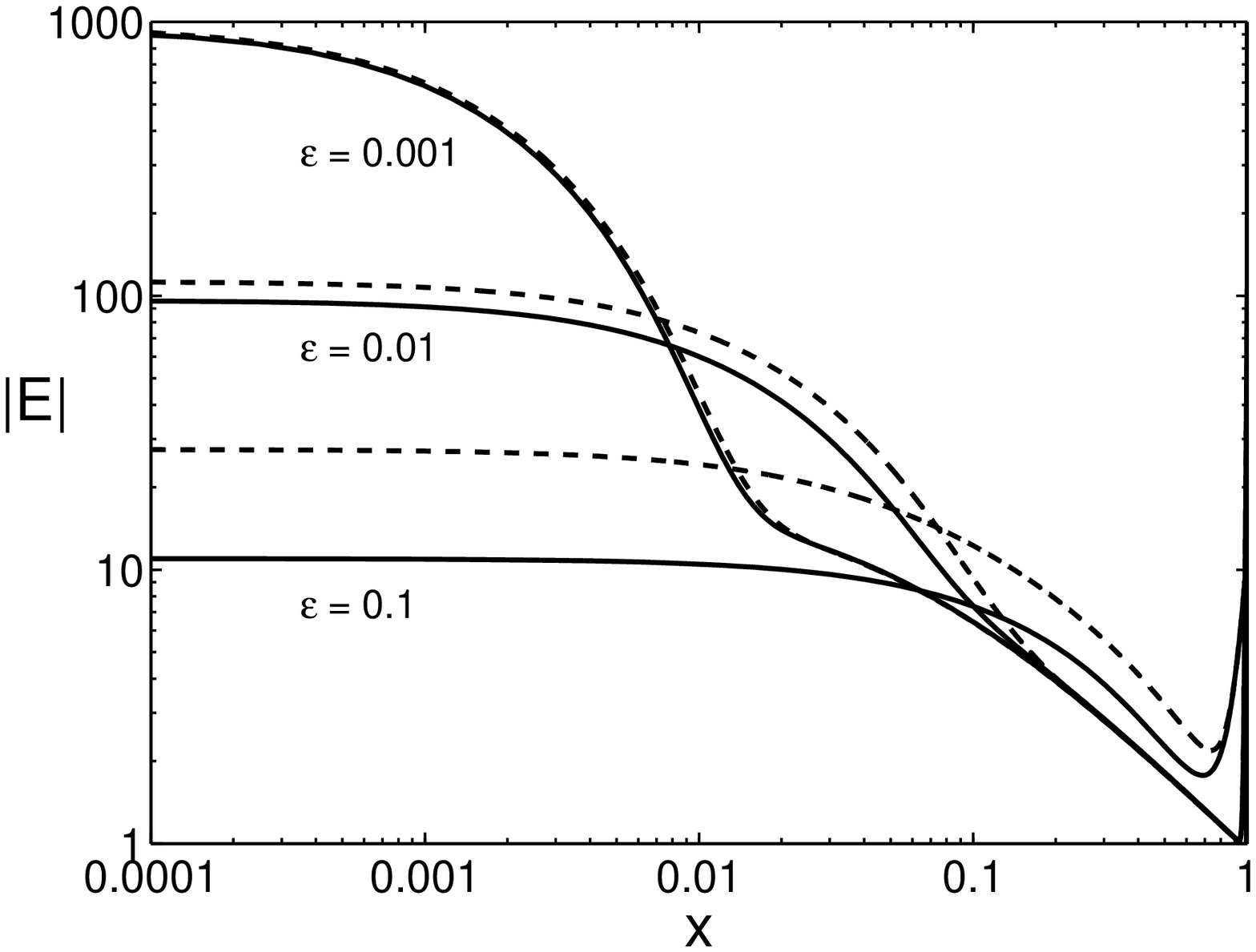}}\\
\scalebox{0.35}{\includegraphics{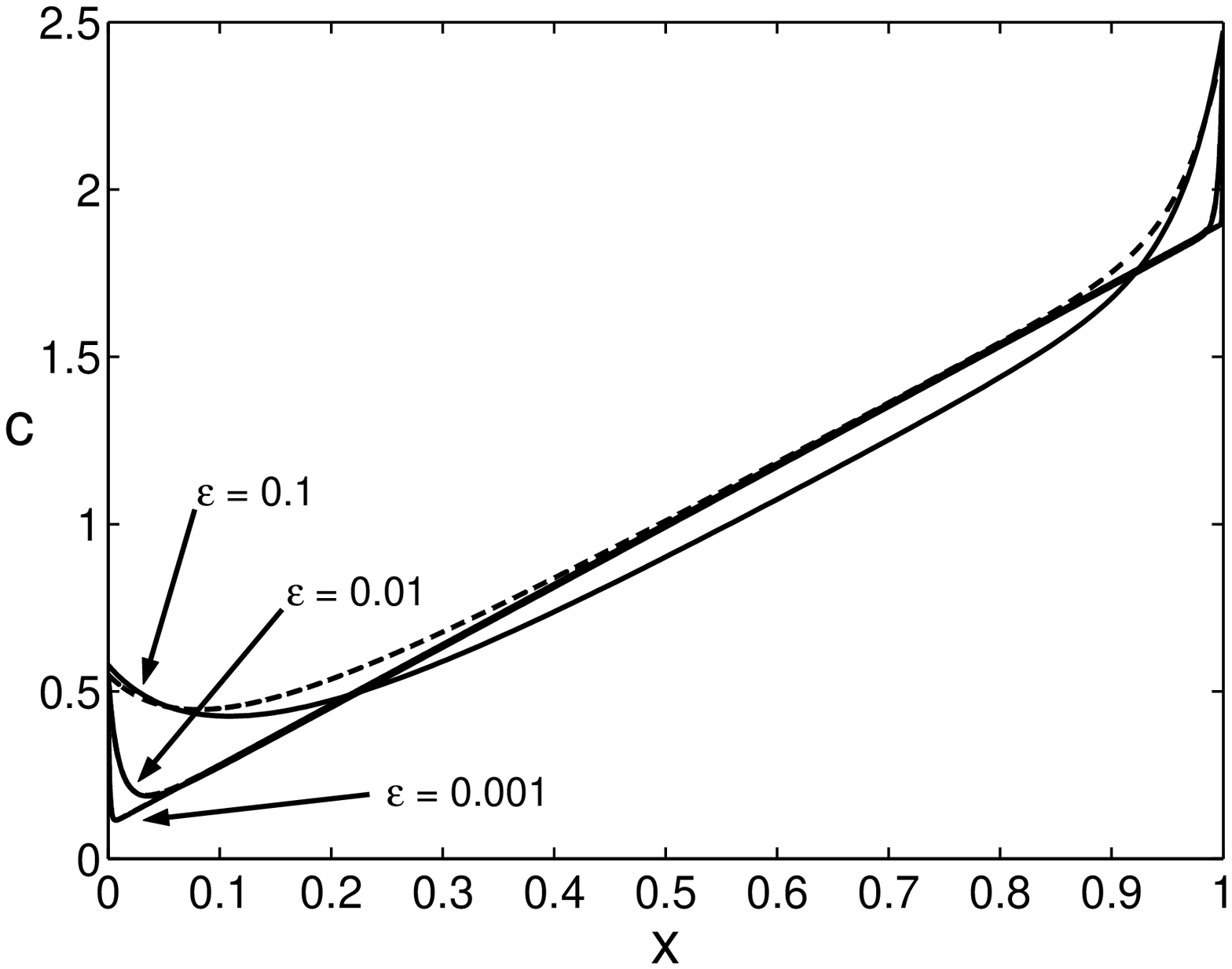}}
\ \nolinebreak 
\scalebox{0.35}{\includegraphics{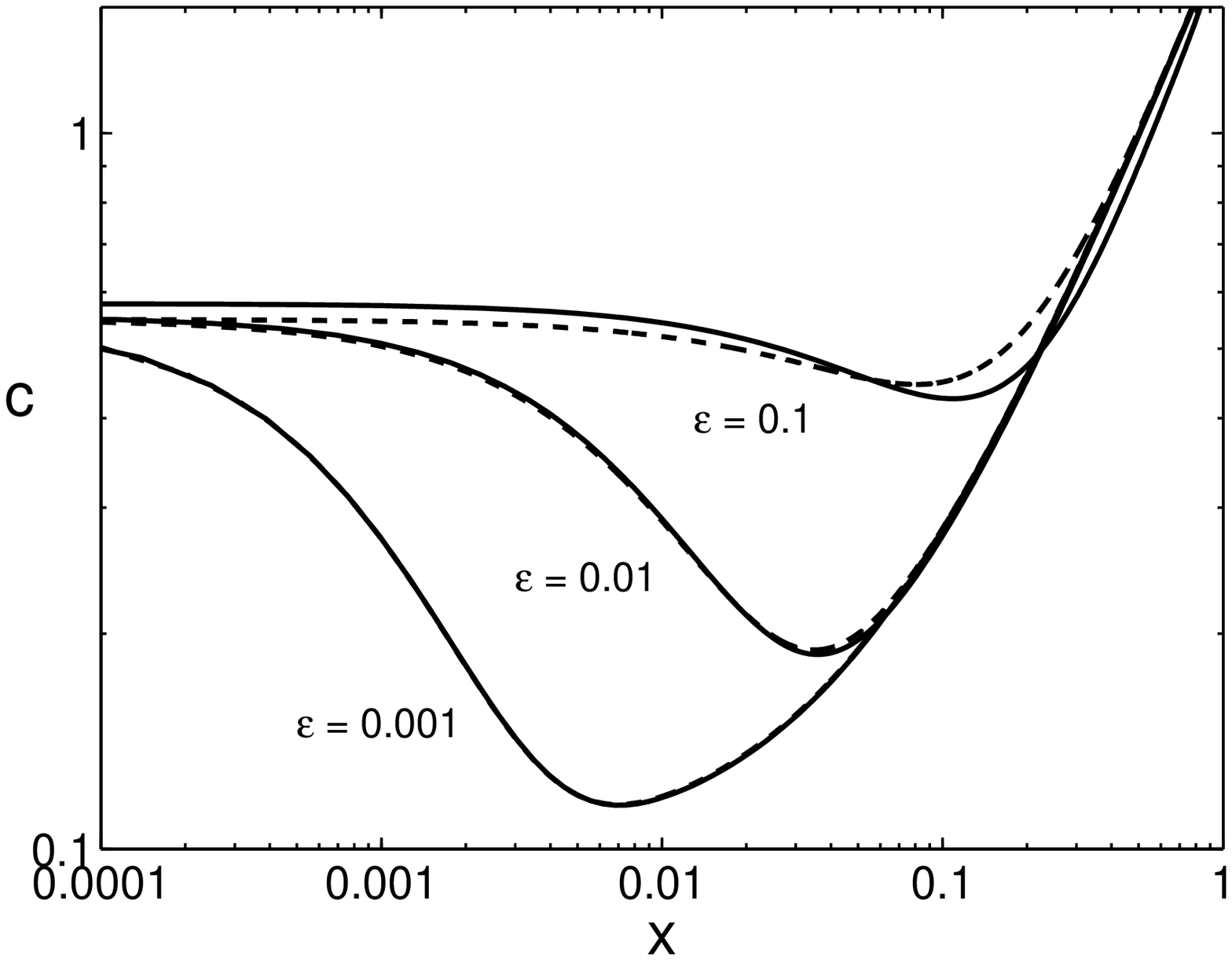}}\\
\scalebox{0.35}{\includegraphics{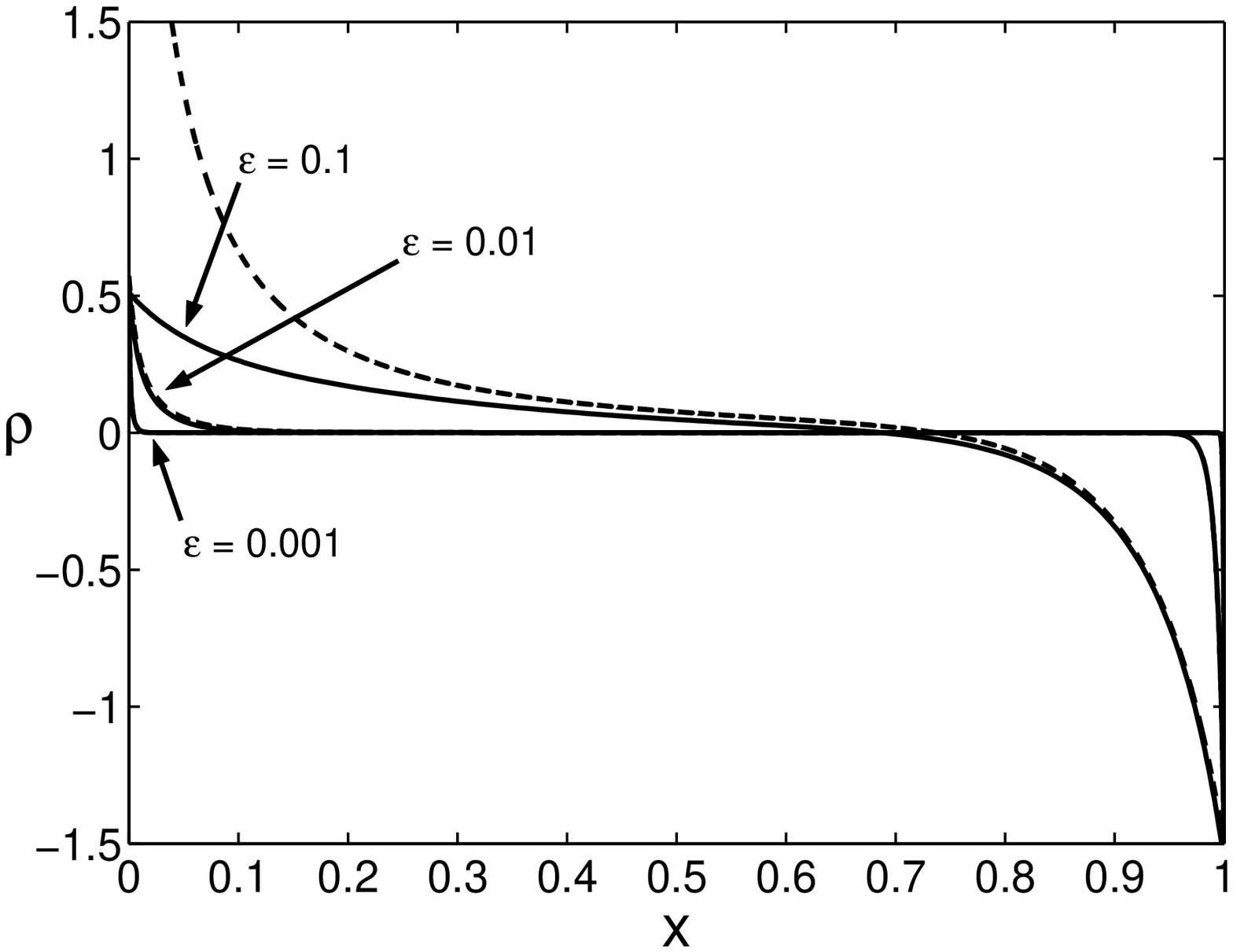}}
\ \nolinebreak 
\scalebox{0.35}{\includegraphics{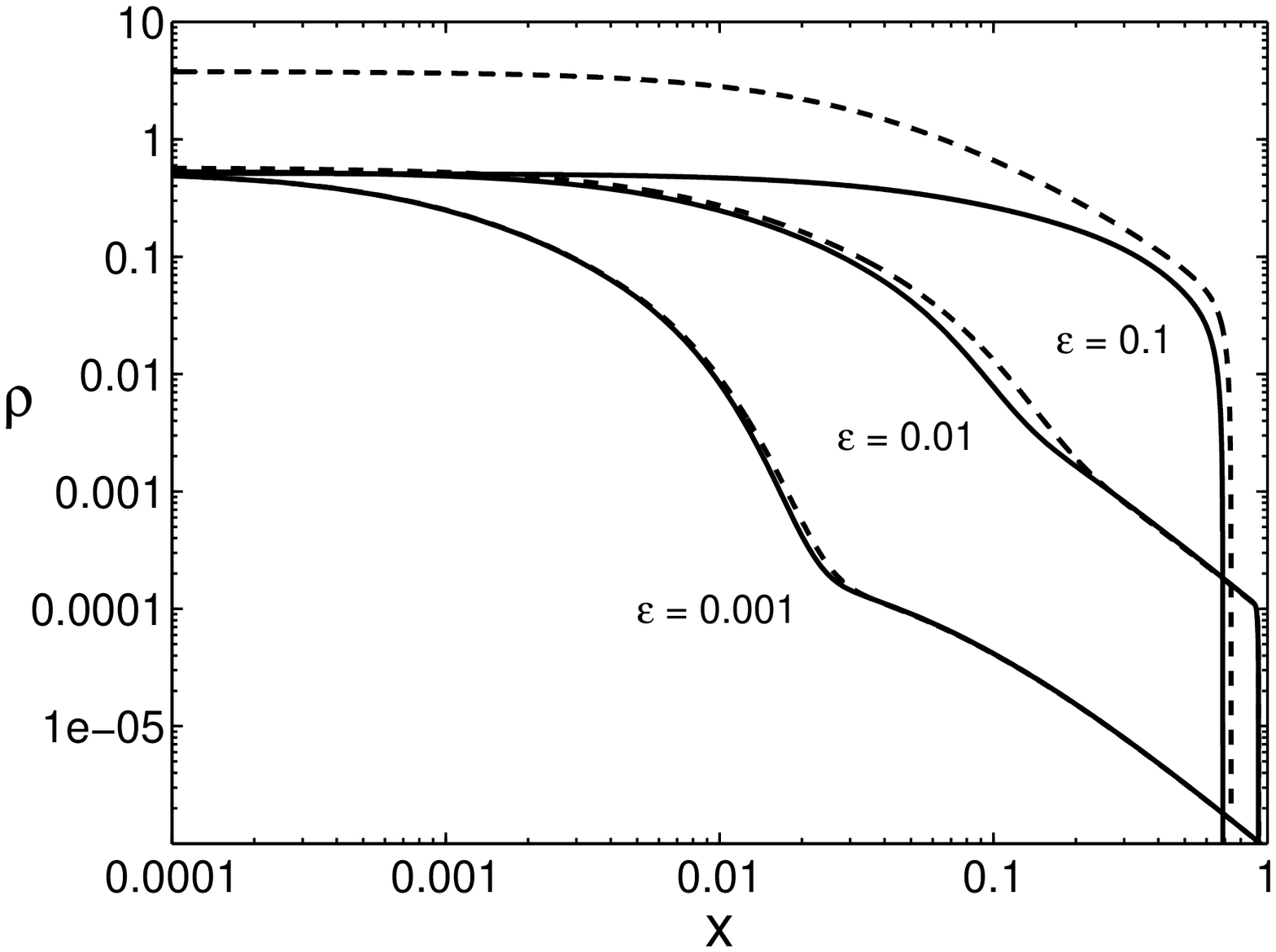}}
\begin{minipage}[h]{5in}
\caption{
\label{figure:full_cell_structure}
Numerical solutions (solid lines) compared with the leading order
uniformly-valid approximations (dashed lines) given by
Eq.~(\ref{eq:cunif})-(\ref{eq:phiunif}) for the dimensionless
potential, $\phi(x)$, electric field, $E(x)$, concentration, $c(x)$,
and charge density, $\rho(x)$ for the case $j=0.9$, $k_c = 10$, $\ir =
10$, $\delta=0$, and $\epsilon=0.001,0.01,0.1$.  Linear scales on the
left show the entire cell, while log scales on the left zoom in on the
cathodic region.  Note that in the concentration and charge density
profiles, the numerical and asymptotic solutions are barely
distinguishable for $\epsilon\leq 0.01$.  }
\end{minipage}
\ec
\end{figure}

Throughout this section, the reader may refer to Figure~
\ref{figure:full_cell_structure}, 
which compares the uniform asymptotic solutions derived below to
numerical solutions at several values of $\epsilon$.  These figures
illustrate the structure of the field variables in the cell as well as
give an indication of the quality of the asymptotic solutions.  The
numerical solutions are obtained by a straightforward iterative
spectral method, described in a companion
paper~\cite{chu2004}.

\subsection{Electroneutrality in the Bulk Solution \label{sec:bulk_soln_below_lim_cur}} 
The most fundamental approximation in electrochemistry is that of bulk
electroneutrality~\cite{newman_book}.  As first emphasized by Newman
\cite{newman1965}, however, this does not mean that the charge
density is vanishing or unimportant, but rather that over macroscopic
distances the charge density is small compared to the total
concentration, $|C_+ - C_-| \ll C_+ + C_-$, or, in our dimensionless
notation, $|\rho| \ll c$. Mathematically, the ``macroscopic limit"
corresponds to the limit $\epsilon = \lambda_D/L
\rightarrow 0$.  The electroneutral solution is just the leading order 
solution when asymptotic series of the form 
$f(x) = f^{(0)}(x) + \epsilon f^{(1)}(x) + \epsilon^2 f^{(2)}(x) + \ldots$,
are substituted for the field variables in 
Eqs.~(\ref{eq:cphieq})-(\ref{eq:phieq}).

Carrying out these substitutions and collecting terms with 
like powers of $\epsilon$, we obtain a hierarchy of differential equations 
for the expansion functions. At $O(1)$ we have,
\bea
\frac{d\bar{c}^{(0)}}{dx} = 2\i \ \ , \ \ 
-\bar{c}^{(0)} \bar{E}^{(0)} = 2\i  \ \ , \ \ 
\bar{\rho}^{(0)} = 0 
\eea
where the bar accent indicates that these expansions are valid 
in the ``bulk region'' 
$\epsilon \ll x \ll 1-\epsilon$ (or $\lambda_D \ll X \ll L-\lambda_D$).
Integating these equations, we obtain the leading order bulk solution:
\bea
\cb^{(0)}(x) &=& c_o + 2\i x  \\ 
\Eb^{(0)}(x) &=& \frac{-1}{x + c_o/2\i} \\
\phib^{(0)}(x) &=& \phi_o + \log\left(1 + \frac{2\i x}{c_o}\right)
\label{eq:bulk_solns}
\eea
where the constants of integration $c_o$ and $\phi_o$ are the values
of the bulk concentration and potential extrapolated to the cathode
surface at $x=0$.  Note that, despite quasi-electroneutrality, the
electrostatic potential does not satisfy Laplace's equation at leading
order in the bulk, as emphasized by Levich \cite{levich_book} and
Newman~\cite{newman_book}.  Noting the presence of $\epsilon^2$ in
Eq.~(\ref{eq:phieq}) for the dimensionless potential, it is clear that
a negligible charge density, $\rho = O(\epsilon^2)$, is perfectly
consistent with a nonvanishing Laplacian of the potential. More
precisely, we have,
\beq
\rhob^{(2)}(x)  = \frac{d^2\phib^{(0)}}{dx^2}(x) = \frac{1}{(x +
c_o/2\i)^2},
\label{eq:rhob2}
\eeq
at $O(\epsilon^2)$ in Eq.~(\ref{eq:phieq}).

The integral constaint, Eq.~(\ref{eq:cint}), can be used to evaluate
the constant $c_o$. If we assume that $c_-$ in the boundary layers
does not diverge as $\epsilon \rightarrow 0$, then they only
contribute $O(\epsilon)$ to the total anion number. Therefore,
\beq
1 = \int_0^1 c_-(x)dx = \int_0^1 \bar{c}^{(0)}(x) dx + O(\epsilon) = c_o +
\i + O(\epsilon) ,
\eeq
which implies that $c_o = 1-\i$.  

At leading order in the bulk, we have recovered the classical theory
dating back a century to Nernst \cite{nernst1904,brunner1904,brunner1907}.
The solution is electrically neutral
with a linear concentration profile whose slope is proportional to the
current. This approximation leads to one of the fundamental concepts in
electrochemistry, that there exists a ``limiting current'', $\i = 1$,
or 
\begin{equation}
\I = \ID = \frac{4zFD_+C_{ref}A}{L} ,
\end{equation}
corresponding to zero concentration at the cathode, $c_o = 1 - \i =
0$. The current is limited by the maximum rate of mass transfer
allowed by diffusion, and larger currents would lead to unphysical and
mathematically inconsistent negative concentrations (see Appendix
\ref{appendix:A}).

Examination of the electric field exposes the same limitation on the
current: a singularity exists at $\i = 1$ that blocks larger currents
from being attained.  The leading order bulk approximation to the
electric field, $\bar{E}^{(0)} = 1/[x + (1-\i)/2\i]$, diverges near
the cathode like $1/x$ in the limit $\i \rightarrow 1$.  This would
imply that the cell voltage $v$ (calculated below) diverges as $\i
\rightarrow 1$, thus providing a satisfactory theory
of the limiting current since an infinite voltage would be necessary
to exceed (or even attain) it. Unfortunately, this classical picture
due to Nernst~\cite{nernst1904}, based on passing to the {\it
singular} limit $\epsilon = 0$, is not valid for any {\it finite} value
of $\epsilon$ because the solution is, in general, unable to satisfy
all of the boundary conditions.

\subsection{Diffuse Charge Layers in Thermal Equilibrium} We now derive the leading
order description of the boundary layers in the standard way
\cite{newman1965,barcilon1997,bard_book,newman_book},
using Eqs.~(\ref{eq:cphieq})--(\ref{eq:rhophieq}).  
The singular perturbation in Eq.~(\ref{eq:phieq})
can be eliminated with the rescaling $y = x/\epsilon$ indicating that 
the boundary layer at $x=0$ has a width $O(\epsilon)$.  
In terms of this inner variable, the governing equations in the
cathode boundary layer are
\bea
\frac{dc}{dy} + \rho \frac{d\phi}{dy} &=& 2 \i \epsilon \label{eq:gc_eqn_c} \\
\frac{d\rho}{dy} + c \frac{d\phi}{dy} &=& 2 \i \epsilon \label{eq:gc_eqn_rho} \\
- \frac{d^2\phi}{dy^2} &=& \rho , \label{eq:gc_eqn_poisson}
\eea
where $\epsilon$ now appears as a regular perturbation since
solutions satisfying the cathode boundary conditions and the matching
conditions still exist when $\epsilon=0$. 
At the anode, the appropriate inner variable is $y = (1-x)/\epsilon$, and 
the equations are the same as above except that $\i$ is replaced with $-\i$, 
since current is leaving the anode layer, while it entering the cathode 
boundary layer.

Expanding the cathode boundary layer fields (indicated by the check accent)
as asymptotic series in powers of $\epsilon$, we obtain at leading order,
\bea
\frac{d\cc^{(0)}}{dy} + \rhoc^{(0)} \frac{d\phic^{(0)}}{dy} = 0 \\
\frac{d\rhoc^{(0)}}{dy} + \cc^{(0)} \frac{d\phic^{(0)}}{dy} = 0 .
\eea
Using Eq.~(\ref{eq:c_and_rho_def}) to rewrite these equations in terms of 
$c_+$ and $c_-$,
we find that the flux of each ionic species in the boundary layer
is zero at leading order: 
\beq
\frac{d\cc_\pm^{(0)}}{dy} \pm \cc_\pm^{(0)} \frac{d\phic^{(0)}}{dy} =
0, \label{eq:zeroflux}.
\eeq
While this equation appears to contradict the fact that the current is 
nonzero, the  paradox is resolved in the same way as 
electroneutrality is reconciled with a non-harmonic potential in the bulk 
region: tiny fluctuations about the boundary layer equilibrium 
concentration profiles at $O(\epsilon)$ are amplified by a scaling 
factor of $1/\epsilon$ to sustain the $O(1)$ current.  Thus, the leading
order contribution to the current in the boundary layer is 
\beq
\epsilon \left ( \frac{d\cc_+^{(1)}}{dx} + \cc_+^{(1)} \frac{d\phic^{(0)}}{dx} 
+ \cc_+^{(0)} \frac{d\phic^{(1)}}{dx} \right ) = 
\frac{d\cc_+^{(1)}}{dy} + \cc_+^{(1)} \frac{d\phic^{(0)}}{dy} 
+ \cc_+^{(0)} \frac{d\phic^{(1)}}{dy} = 4 \i .
\eeq
Integrating Eq.~(\ref{eq:zeroflux}) and matching with the bulk,
we find that the leading order ionic concentrations are Boltzmann 
equilibrium distributions\footnote{The expresssion for energy in 
the Boltzmann equilibrium distribution includes only the energy due 
to electrostatic interactions.
``Chemical'' contributions to the energy are neglected.}: 
\beq
\cc_\pm^{(0)}(y) = c_o e^{\pm[\phi_o - \phic^{(0)}(y)]}
\label{eq:boltz_equilibrium_conc} 
\eeq
where $c_o = 1-j$ and $\phi_o = \phib^{(0)}(0)$ are obtained by matching 
with the solution in the bulk.
Note that the Boltzmann distribution arises not from an assumption of 
thermal equilibrium in the boundary layer but as the leading order 
concentration distribution, even in the presence of a non-negligible 
$O(1)$ current.

The general leading-order solution was first derived by
Gouy~\cite{gouy1910} and Chapman~\cite{chapman1913} and appears
in numerous books~\cite{newman_book,bard_book,hunter_book,lyklema_book} and
recent papers~\cite{barcilon1997,bonnefont2001}:
\bea
\cc^{(0)}(y) & = & c_o \cosh[\phi_o - \phic^{(0)}(y)]
\label{eq:boltz_c} \\
\rhoc^{(0)}(y) & = & c_o \sinh[\phi_o - \phic^{(0)}(y)]
\label{eq:boltz_rho}  \\
\frac{d\phic^{(0)}}{dy} & = & 2 \sqrt{c_o} \sinh\left(\frac
{ \phi_o - \phic^{(0)}(y)}{2} \right) \label{eq:E_gc} \\
\phic^{(0)}(y) &=& \phi_o + 4 \tanh^{-1}\left(\gamma_o
e^{-\sqrt{c_o}y}\right) , \label{eq:potgc}
\eea
where $\gamma_o \equiv \tanh(\zeta_o/4)$ and $\zeta_o \equiv
\phic^{(0)}(0) - \phi_o$ is the leading order ``zeta potential''
across the cathodic diffuse layer, which plays a central role in
electrokinetic phenomena~\cite{hunter_book,lyklema_book}.  Note that 
the magnitude of the diffuse layer electric field 
scales as $1/\epsilon$ as illustrated in 
Figure~\ref{figure:full_cell_structure}.

The value of $\phic^{(0)}(0)$ hidden in the zeta potential
$\zeta_o$ is determined by the Stern boundary condition,
Eq.~(\ref{eq:potbc0}). If $\delta = 0$ (Gouy-Chapman model), then
$\phic^{(0)}(0)=0$, or $\zeta_o = -\phi_o$, which means that the
entire voltage drop $\phi_o$ across the cathodic double layer occurs
in the diffuse layer. If $\delta = \infty$ (Helmholtz model), then
$\phic^{(0)}(0) = \phi_o$, or $\zeta_o = 0$, in which case the
Stern layer carries all the double layer voltage. For finite $\delta >
0$ (Stern model), $\zeta_o$ is obtained in terms of $\phi_o$ by
solving a trancendental algebraic equation,
\beq
-\zeta_o = 2 \delta \sqrt{c_o} \sinh(\zeta_o/2) + \phi_o ,
\label{eq:sterneqc}
\eeq
which can be linearized about the two limiting cases and solved for
$\zeta_o$,  
\beq
- \zeta_o \sim \left\{ \begin{array}{ll} 
\phi_o - 2 \delta \sqrt{c_o} \sinh(\phi_o/2)  &
 \mbox{ if } \delta  \ll \phi_o/2\sqrt{c_o}\sinh(\phi_o/2) \\
  \phi_o/\delta \sqrt{c_o} &
 \mbox{ if } \delta \gg \phi_o/2\sqrt{c_o}
\end{array} \right.
\eeq 
Note that if $\phi_o \ll 1$, then 
$-\zeta_o \approx \frac{\phi_o}{1 + \delta \sqrt{c_o}}$
is a reasonable approximation for any value of $\delta \geq 0$.
Finally, we solve for $\phi_o$ by applying the
Butler-Volmer rate equation, Eq.~(\ref{eq:potbvbc0}), which
yields a transcendental algebraic equation for 
$\phi_o$:
\beq
k_c c_o e^{-\zeta_o + \alpha_c(\zeta_o + \phi_o)} - \ir
e^{-\alpha_a(\zeta_o + \phi_o)}  =  \i.
\label{eq:bveqc}
\eeq
Simultaneously solving the pair of equations (\ref{eq:sterneqc}) and
(\ref{eq:bveqc}) exactly is not possible in general, but below we will
analyze various limiting cases.

In the anodic boundary layer, we find the same set of equations as
Eqs.~(\ref{eq:gc_eqn_c})-(\ref{eq:gc_eqn_poisson}) except 
that $\i$ is replaced by $-\i$. 
Therefore, since the fields do not depend on $\i$ at leading order, 
the anodic boundary layer has the same structure but with different 
constants of integration. Thus, we find that the leading order 
description of the anodic boundary layer is given by 
Eqs~(\ref{eq:boltz_c})-(\ref{eq:potgc}) with 
$c_o$, $\phi_o$, $\gamma_o$, and $\zeta_o$ 
replaced by different constants
$c_1$, $\phi_1$, $\gamma_1$, and $\zeta_1$, respectively.
Moreover, it is straightforward to show that 
$c_1 = \cb^{(0)}(1) = 1+\i$ 
and 
$\phi_1 = \phi_o + \log\left(\frac{1+\i}{1-\i}\right)$.


The leading-order anodic zeta potential, $\zeta_1$, and potential drop
across the entire anodic double layer, $v - \phi_1$,
are found by solving another pair of trancendental algebraic equation 
resulting from the anode Stern and Butler-Volmer boundary conditions,
Eqs.~(\ref{eq:potbc1}) and (\ref{eq:potbvbc1}),
\bea
-\zeta_1 & = & 2\delta\sqrt{c_1} \sinh(\zeta_1/2) 
+ \phi_1 - v
\label{eq:sterneqa} \\
\i & = & - k_c c_1 e^{-\zeta_1 + \alpha_c(\zeta_1 + \phi_1 - v)} + \ir
e^{-\alpha_a(\zeta_1 + \phi_1 - v)}. \label{eq:bveqa}
\eea
As before, the Gouy-Chapman and Helmholtz limits are
$\zeta_1 = v - \phi_1$ and $\zeta_1 = 0$,
respectively, and for small voltages (or currents) the approximation
$\zeta_1 \approx (v-\phi_1)/(1+\delta\sqrt{c_1})$
is valid for all $\delta \geq 0$.

\subsection{Leading Order Uniformly-valid Approximations} 
We obtain asymptotic approximations that are uniformly valid across the 
cell by adding the bulk and boundary layer approximations and
subtracting the overlapping parts: 
\bea
c(x) & = & [\cc^{(0)}(x/\epsilon) - c_o] + \cb^{(0)}(x) + 
	[\ch^{(0)}((1-x)/\epsilon) - c_1] + O(\epsilon) ,
	\label{eq:cunif} \\ 
\rho(x) & = & \rhoc^{(0)}(x/\epsilon) + \epsilon^2 \rhob^{(2)}(x) + 
	\rhoh^{(0)}((1-x)/\epsilon)+ O(\epsilon) , \label{eq:rhounif}\\ 
E(x) & = & \frac{1}{\epsilon} \frac{d\phic^{(0)}}{dy}(x/\epsilon) +
	\frac{d\phib^{(0)}}{dx}(x) - 
	\frac{1}{\epsilon} \frac{d\phih^{(0)}}{dy}((1-x)/\epsilon) + 
	O(\epsilon),
	\label{eq:Eunif}  \\
\phi(x) & = & [\phic^{(0)}(x/\epsilon) - \phi_o] + \phib^{(0)}(x) + 
	[\phih^{(0)}((1-x)/\epsilon) - \phi_1] + O(\epsilon)
	\label{eq:phiunif} .
\eea
Note that we have kept the $O(\epsilon^2)$ term in the charge density
since it is the leading order contribution in the bulk region and is
easily computed from Eq.~(\ref{eq:rhob2}).  As shown in
Figure~\ref{figure:full_cell_structure}, the leading-order uniformly
valid solutions are very accurate for $\epsilon \leq 0.01$ (or $L \geq
100 \lambda_D$) and reasonably good for $\epsilon=0.1$.  Since
higher-order terms are not analytically tractable, it seems numerical
solutions must suffice for nanolayers, where $\epsilon \approx 1$, or
else other limits of various parameters must be considered, as below.

The discrepancy in electric potential profile at large $\epsilon$ in
Figure~\ref{figure:full_cell_structure} is particularly interesting
because it arises from a constraint on the total potential drop across
the cell.  To understand the origin of this voltage constraint, recall
that the total cell voltage is determined by the current density
flowing through the cell (via the voltage-current relationship).
While $\epsilon$ is technically a parameter in the voltage-current
relationship, a leading-order analysis does not capture the $\epsilon$
dependence.  Thus, the leading-order cell voltage must be the same for
all $\epsilon$ which is what we observe in figure
\ref{figure:full_cell_structure}.  A close examination of the
potential and electric field profiles reveals that most of the error
in the asymptotic solution for the potential comes from an over
prediction of the electric field strength (and therefore the potential
drop) in cathode region.

\section{Polarographic Curves for Thin Double Layers, $\epsilon
\rightarrow 0$} The relationship between current 
and cell voltage is of primary importance in the study of any
electrochemical system, so we now use the results from the previous
section to calculate theoretical polarographic curves in several
physically relevant regimes.  We focus on the effects of the Stern
capacitance and the reaction rate constants through the dimensionless
parameters, $\delta$, $k_c$ and $\ir$, with $\alpha_c = \alpha_a=1/2$.
For a fixed voltage, the mathematical results are valid in the
asymptotic limit of thin double layers, $\epsilon
\rightarrow 0$. 

\subsection{ Exact Results at Leading Order}
Using the uniformly valid approximation Eq.~(\ref{eq:phiunif}), we can
write the leading order approximation for the cell voltage as
\beq
v = \phi_o + 2 \tanh^{-1}(\i) + (v-\phi_1) \label{eq:vi_gen_form}.
\eeq
We can interpret this expression as a decomposition of the cell
voltage into the potential drop across the cathode, bulk, and anode
layers respectively.  Note the divergence in the bulk contribution to
the cell voltage as $\i \rightarrow 1$, which we expect from our
earlier analysis.  In the next section, we explore analytic solutions
for several limiting cases and compare them to exact solutions given
by Eq.~(\ref{eq:vi_gen_form}) with the leading-order cathode and anode
diffuse layer potential drops determined implicitly by
Eqns.~(\ref{eq:sterneqc}), (\ref{eq:bveqc}), (\ref{eq:sterneqa}). To
make plots in our figures, we use Newton iteration to solve for
$\phi_o$ and $v-\phi_1$ in this algebraic system.

\subsection{ Cell Resistance at Low Current }
Given the common practice of using linear circuit models to describe
electrochemical systems~\cite{macdonald1990,geddes1997,bazant2004}, it
is important to consider the low-current regime, where the cell acts
as a simple resistance, $R = V/\I$.  First, we compute the potential
drop across the double layers.  Since the procedure is almost
identical for the two boundary layers, we focus on the calculation for
the cathode.  By writing the boundary conditions
Eq.~(\ref{eq:potbvbc0}) in the standard Butler-Volmer form involving
the exchange current and surface overpotential
\cite{bard_book,newman_book}:
\beq
  \i = \i_o^{c} \left ( e^{-\alpha_c \eta_s^{c}} - e^{\alpha_a \eta_s^{c}} 
              \right )
  \label{eq:bvc_standard}
\eeq
where 
$\i_o^c = \left ( k_c c_o e^{-\zeta_o} \right )^{\alpha_a} \ir^{\alpha_c}$
and $\eta_s^c = \Delta \phi_S - \Delta \phi_S^{eq}$ 
are the cathode exchange current and surface overpotential, respectively.  
Note that the exchange current contains the Frumkin correction through
the factor $e^{-\zeta_o}$ \cite{bard_book}.  For low current densities, 
we expect the surface overpotential to be small, so we can linearize
this equation to obtain
\beq
  \i \sim -\i_o^{c} \eta_s^{c} 
  \label{eq:bvc_low_cur}
\eeq
where we have used the fact that $\alpha_c + \alpha_a = 1$.
Rewriting this equation in terms of $\phic(0)$, we find that
\beq
  \phic(0) \sim \frac{\i}{\i_o^c} + \phic_{eq}(0),
  \label{eq:phi0_low_cur}
\eeq 
where $\phic_{eq}(0)$ is the value of $\phic(0)$ calculated from the 
cathode Butler-Volmer rate equation when there is no current flowing 
through the electrode. 
The zeta potential $\zeta_o$ in the formula for the exchange current
is determined by combining Eq.~(\ref{eq:phi0_low_cur}) with
(\ref{eq:sterneqc}) to obtain a single equation for $\zeta_o$:
\beq
  -2 \delta \sqrt{c_o} \sinh \left( \zeta_o /2 \right )
  \sim \frac{\i}{\left ( k_c c_o e^{-\zeta_o} \right )^{\alpha_a} \ir^{\alpha_c}}
  + \log \left ( \frac{\ir}{k_c c_o} \right ) + \zeta_o.
  \label{eq:sternbc0_low_cur}
\eeq
Finally, to compute the total double layer potential drop 
we add the potential drop across the diffuse layer to $\phic(0)$:
\beq
  \phi_o = \phic(0) - \zeta_o \sim  
       \frac{\i}{\i_o^c} + \log \left ( \frac{\ir}{k_c c_o} \right ).
  \label{eq:phi_o_low_cur}
\eeq
A similar calculation at the anode results in 
\beq
  v - \phi_1 \sim \frac{\i}{\i_o^a} + \log \left ( \frac{k_c c_1}{\ir} \right ),
  \label{eq:phi_1_low_cur}
\eeq
where 
$\i_o^a = \left ( k_c c_1 e^{-\zeta_1} \right )^{\alpha_a} \ir^{\alpha_c}$
and $\zeta_1$ is determined by the anode equivalent of 
Eq.~(\ref{eq:sternbc0_low_cur}).

Combining these results with the potential drop across the bulk solution, 
we find that the total cell voltage is given by
\bea
  v(\i) &\sim& 4 \tanh^{-1}(\i) 
            + \frac{\i}{\i_o^c} + \frac{\i}{\i_o^a} \nonumber \\
        &\approx& \i \left (4
                  + \frac{1}{\i_o^c} + \frac{1}{\i_o^a} \right ).
  \label{eq:vi_low_cur} \\
& = & \i\; r \nonumber
\eea
This result gives the dimensionless resistance, $r$, of the
electrochemical thin film as a function of the physical properties of
the electrodes and the electrolyte.  Note that the Stern-layer
capacitance is accounted for implicitly via the calculation of the
electrode zeta potentials.

\begin{figure}[htb]
\bc
\scalebox{0.5}{\includegraphics{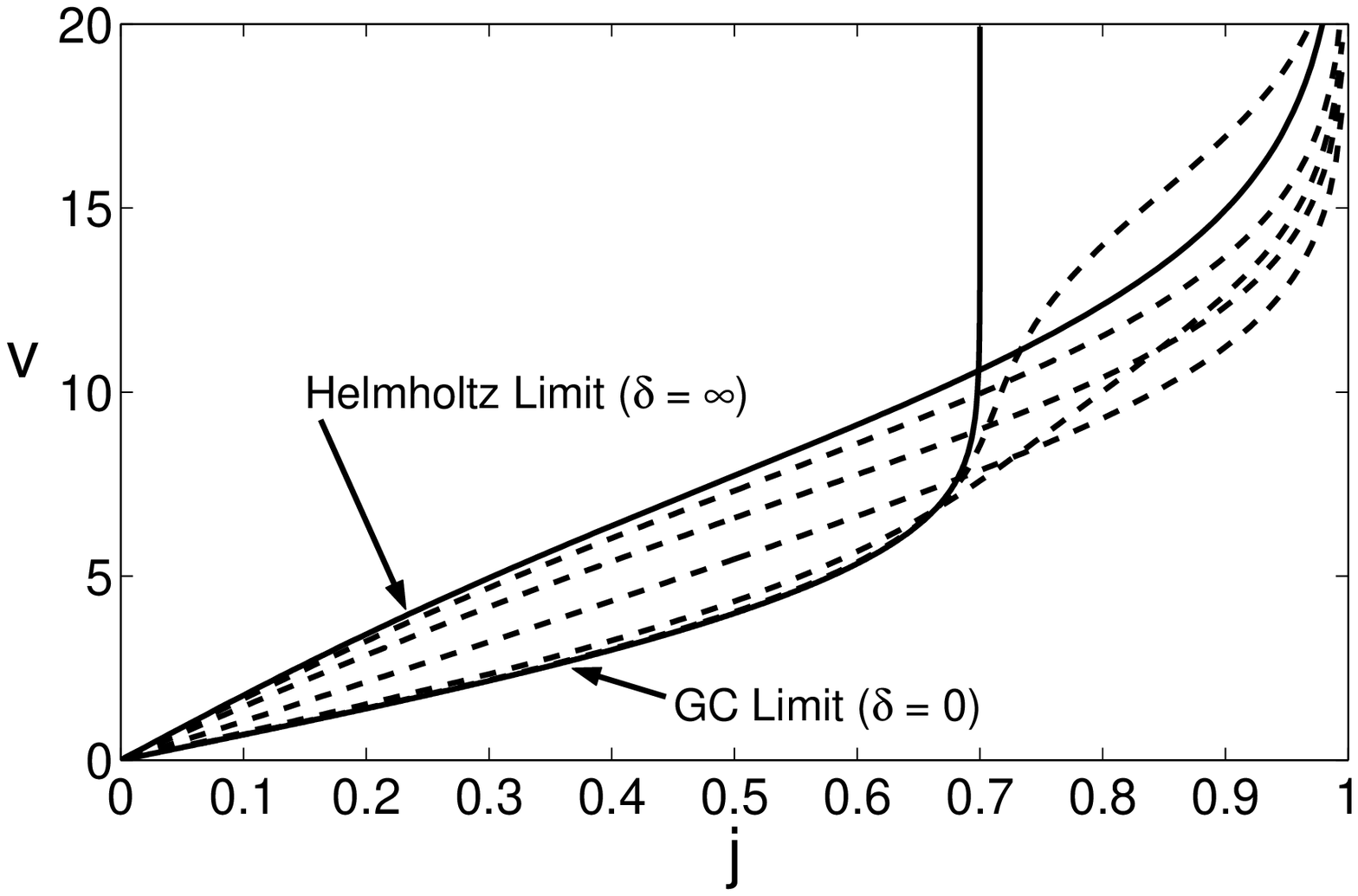}}\\
\scalebox{0.5}{\includegraphics{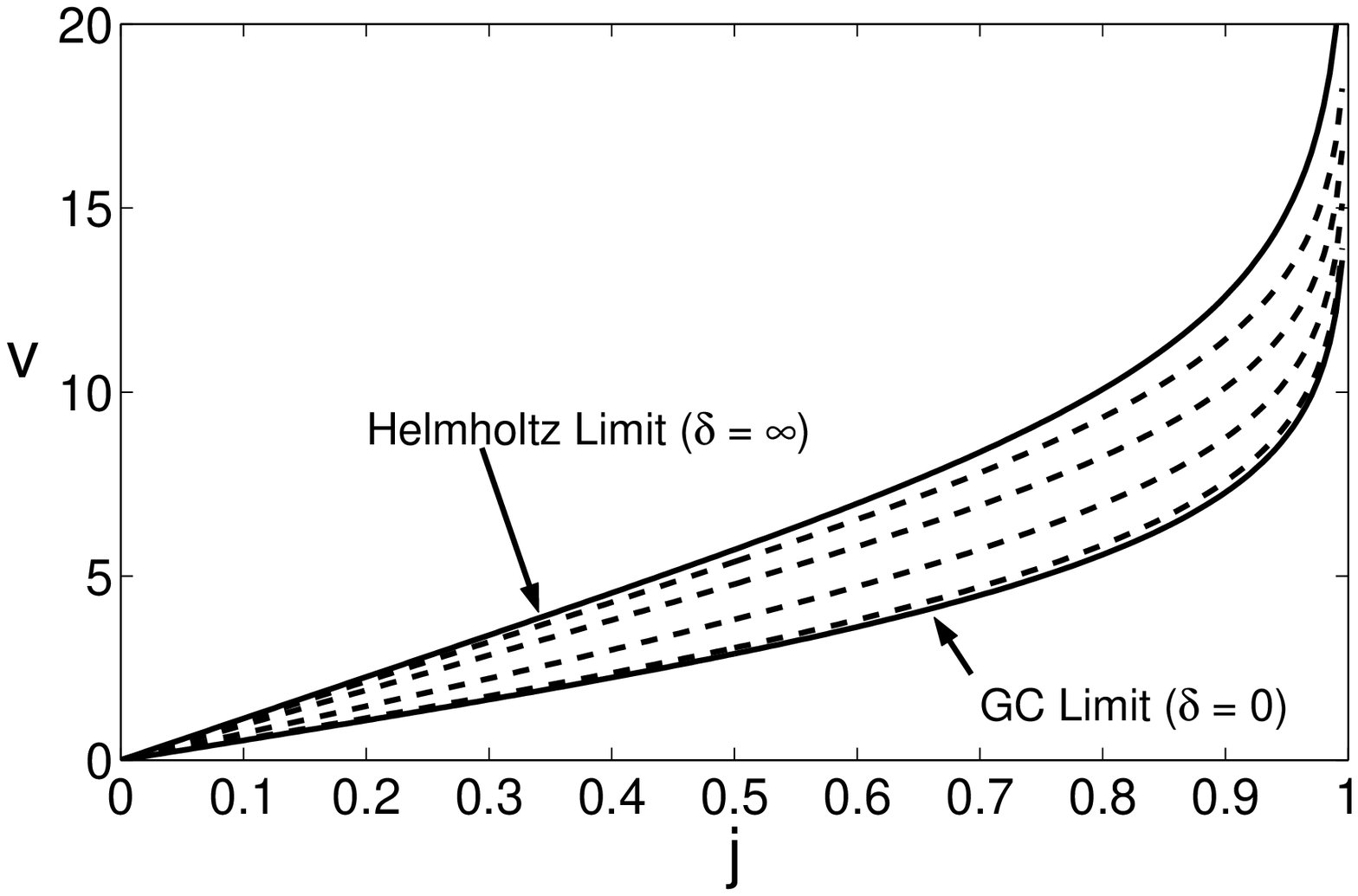}}
\begin{minipage}[h]{5in}
\caption{
\label{figure:jv_rxn_vs_diffusion_lim}
Exact polarographic curves (dashed lines) for varying $\delta$ values
compared to polarographic curves for the Gouy-Chapman ($\delta=0$) and
Helmholtz ($\delta=\infty$) limits (solid lines). Top: a
reaction-limited cell ($\ir<1$) with physical parameters, $k_c = 0.03,
\ir = 0.7$.  Notice that above the reaction-limited current density,
$\ir$, the highest cell voltages occur for $\delta$ values near $0$.
Bottom: a diffusion-limited cell ($\ir > 1$) with physical
parameters: $k_c = 0.05, \ir = 1.5$. In both cases, $\delta$ increases
as the curves move upwards.  }
\end{minipage}
\ec
\end{figure}

\subsection{Simple Analytical Formulae} The exact leading-order
current-voltage relation simplifies considerably in a variety of
physically relevant limits. These approximate formulae provide insight
into the basic physics and may be useful in interpretting experimental
data.

\subsubsection{The Gouy-Chapman Limit $(\delta \rightarrow 0)$} 
In this limit, the capacitance of the diffuse layer of the charged double 
layer is negligible compared to the capacitance of the compact layer.  
As a result, the voltage drop across the diffuse layer accounts for the 
entire potential drop across the charged double layer.
Physically, this limit corresponds to the limits of low ionic concentration
or zero ionic volume \cite{bard_book}.
Since $\zeta_o = -\phi_o$ and $\zeta_1 = v - \phi_1$ 
when $\delta = 0$, the Butler-Volmer rate equations, 
Eqs.~(\ref{eq:bveqc}) and (\ref{eq:bveqa}), reduce to 
\bea
k_c (1-\i) e^{-\zeta_o} - \ir = \i \ \ \textrm{and} \ \ 
- k_c (1+\i) e^{-\zeta_1} + \ir  =  \i.
\eea
Solving for $\zeta_o$ and $\zeta_1$, we find that 
\bea
  \zeta_o = \ln \left ( 1-\i \right ) 
   - \ln \left (\frac{\ir+\i}{k_c} \right )  \ \ \textrm{and} \ \ 
  \zeta_1 = \ln \left ( 1+\i \right ) 
   + \ln \left ( \frac{k_c}{\ir-\i} \right ),
\eea
which can be substituted into Eq.~(\ref{eq:vi_gen_form}) to obtain
\beq
  v(\i) = 4 \tanh^{-1}(\i) + 2 \tanh^{-1} (\i/\ir). \label{eq:vi_gc}
\eeq
Notice that the boundary layers make a nontrivial contribution to the
leading order cell voltage.  The $2 \tanh^{-1}(\i/\ir)$ term is
especially interesting because it indicates the existence of a
reaction-limited current when $\ir < 1$.  In hindsight, it is obvious
that reaction limited currents exist in the Gouy-Chapman limit because
the reaction kinetics at the anode do not permit a current greater
than $\ir$.  We emphasize, however, that the Gouy-Chapman limit is
singular because there is no problem achieving current densities above
$\ir$ for any $\delta > 0$ (see Figure~\ref{figure:jv_rxn_vs_diffusion_lim}).  
For any finite $\delta > 0$, the shift of the anode double-layer potential 
drop to the Stern layer helps the dissolution reaction while suppressing
the deposition reaction which permits the current density
to rise greater than $\ir$.

Note that the cathodic and anodic boundary layers do not evenly
contribute to the cell voltage near the limiting currents.  In a
diffusion-limited cell, the cathodic layer makes the greater
contribution because as $\i \rightarrow 1$, $\zeta_o$ diverges while
$\zeta_1$ approaches a finite limit.  We expect this behavior because
as $\i \rightarrow 1$, the electric field only diverges at $x=0$.
However, when the cell is reaction-limited, the division of cell
voltage between the boundary layers is reversed as $\i$ approaches the
limiting current $\ir$.  Even the voltage drop in the bulk becomes
negligible compared to $\zeta_1$ in the reaction-limited case.  In
this situation, the cell voltage diverges as $\i \rightarrow \ir$
because the only way to achieve a current near $\ir$ is to drastically
reduce the deposition reaction at the anode.  In other words, the
cation concentration at the anode must be made extremely small which
requires a huge anodic zeta potential.

\subsubsection{The Helmholtz Limit $(\delta \rightarrow \infty)$} 
This is the reverse of the Gouy-Chapman limit.  Here, the capacitance
of the compact layer is negligible, so the potential drop across the
double layer resides completely in the compact layer. The Helmholtz
limit holds for concentrated solutions or solvents with low dielectric
constants and other situations where the Debye screening length
becomes negligible~\cite{bard_book}.  It also describes a thick
dielectric or insulating layer on an
electrode~\cite{ajdari2000,bazant2004}.

In the Helmholtz limit, $\zeta_o = 0 = \zeta_1$, so  
the Butler-Volmer rate equations take the form
\bea
k_c (1-\i) e^{\alpha_c \phi_o} - \ir e^{-\alpha_a \phi_o} & = & \i \\
- k_c (1+\i) e^{\alpha_c (\phi_1-v)} + \ir e^{-\alpha_a (\phi_1-v)} & = & \i.
\eea
Solving these equations for $\phi_o$ and $v-\phi_1$ under the assumption of a
symmetric electron-transfer reaction ({\it i.e.} 
$\alpha_c = 1/2 = \alpha_a$) and substituting into the 
formula for the cell voltage, we find that 
\beq
  v(\i) = 6 \tanh^{-1}(\i) 
   + 2 \ln \left ( 
   \frac{\i+\sqrt{\i^2+4 \ir k_c (1-\i)}}{-\i+\sqrt{\i^2+4 \ir k_c (1+\i)}}
  \right ). 
  \label{eq:vi_helmholtz}
\eeq
While this expression appears to be more complicated than the one obtained
for the Gouy-Chapman model, it is not very different when 
$\ir > 1$ as can be seen in Figures~\ref{figure:jv_rxn_vs_diffusion_lim} and 
\ref{figure:fast_rxn_limits}.  In fact, the wide spread in the 
polarographic curves observed in Figure~\ref{figure:jv_rxn_vs_diffusion_lim}
requires that $k_c \ll \ir$; otherwise, all of the curves would be
difficult to distinguish.  Moreover, as we shall see in the next
section, in limit of fast reactions, both models lead to the same
expression for the cell voltage for $\ir > 1$.  On the other hand,
when $\ir < 1$, the two models are qualitatively very different.
While the Gouy-Chapman model gives rise to a reaction-limited current,
the Helmholtz model does not.

\subsubsection{The Fast-Reaction Limit, 
$(\ir \gg 1, (\ir)^{\alpha_a} (k_c)^{\alpha_c} \gg
\i/(1-\i)^{\alpha_a})$ } 
\begin{figure}[htb]
\bc
\scalebox{0.5}{\includegraphics{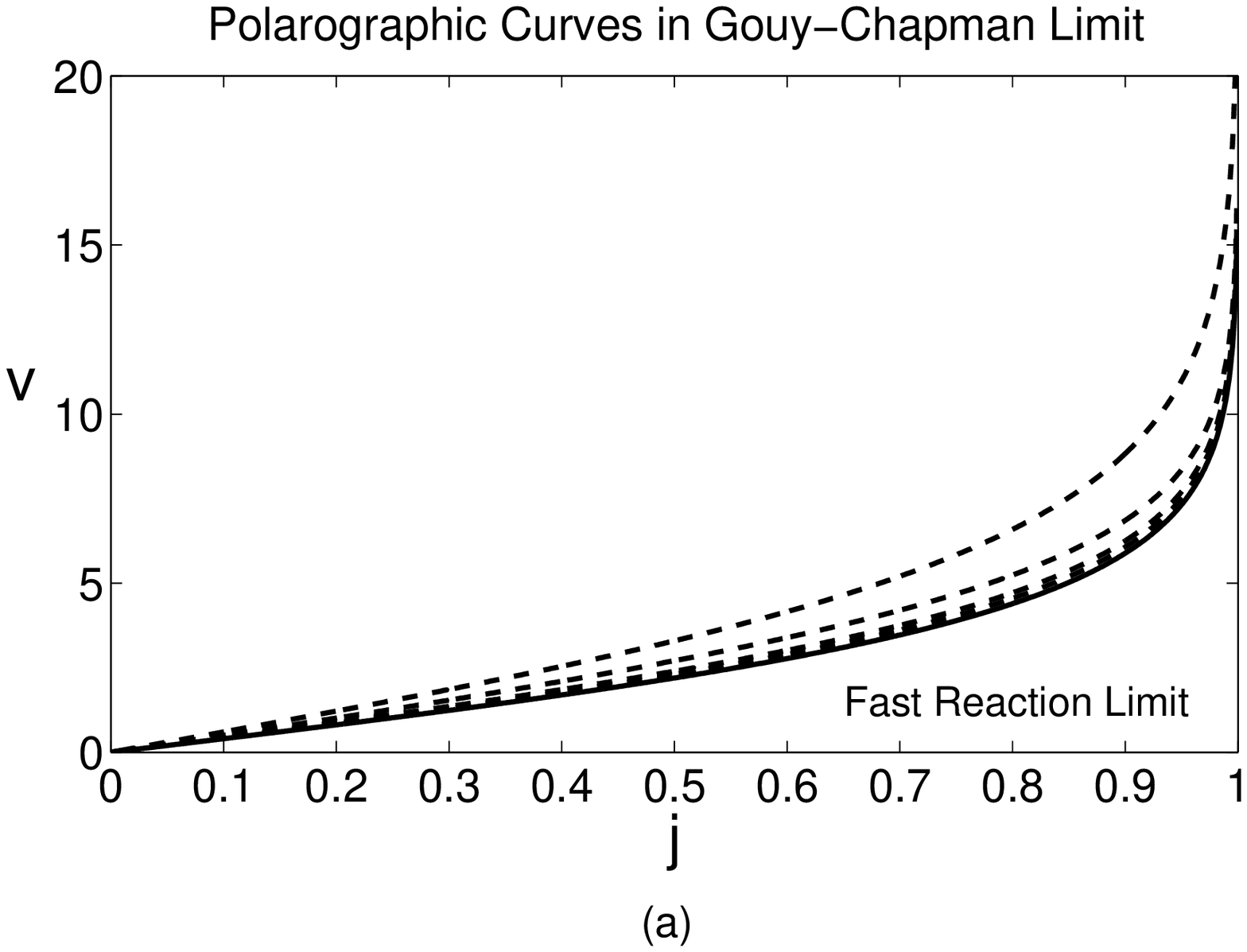}} \\
\scalebox{0.5}{\includegraphics{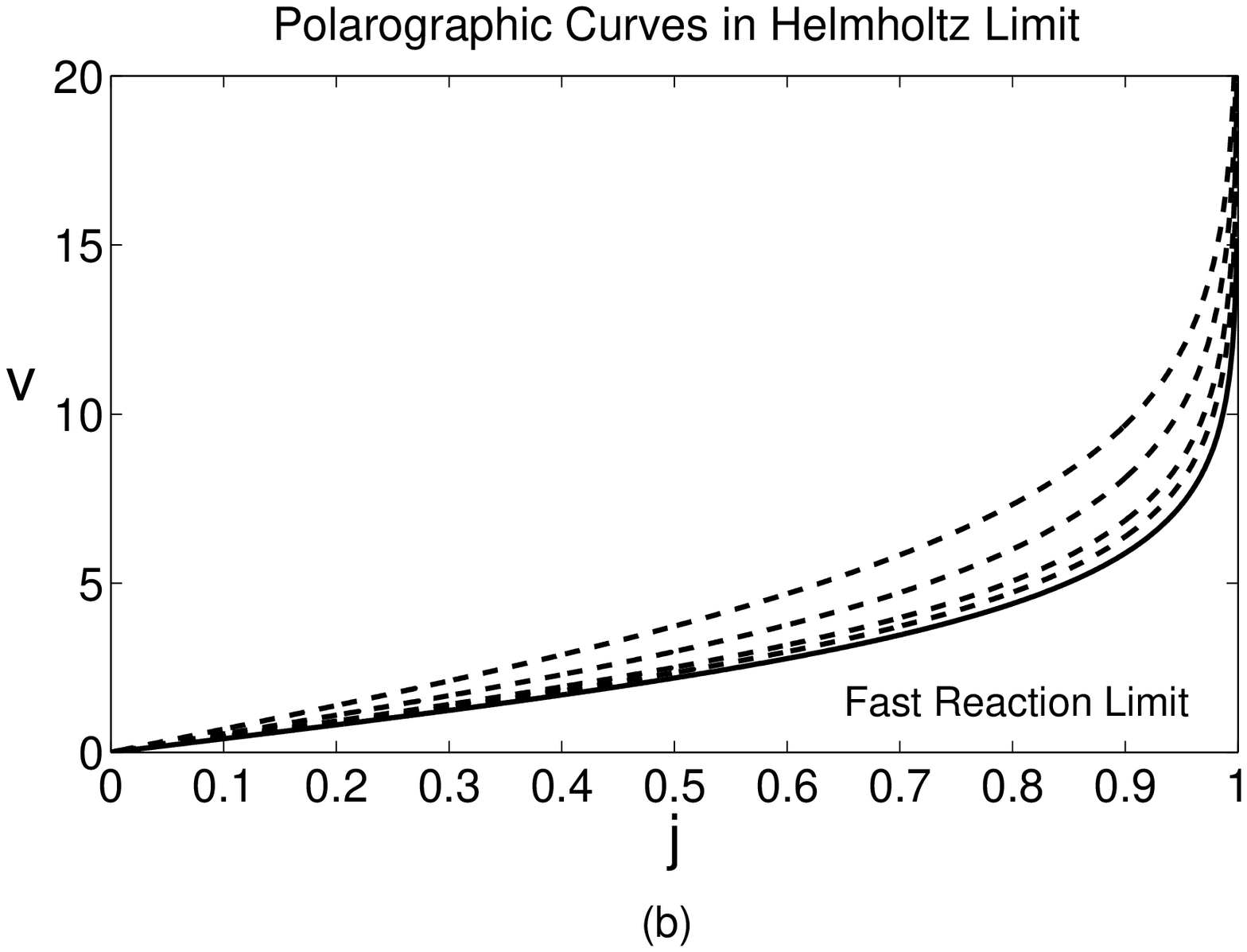}}
\begin{minipage}[h]{5in}
\caption{
\label{figure:fast_rxn_limits}
Polarographic curves in the (a) Gouy-Chapman limit
and (b) Helmholtz limit as the reaction rate 
constants are increased (dashed lines).  
For these plots, the reaction rate constants ($\ir = 1, 2, 5, and 10$) 
increase as the curves shift towards the lower right and are related by 
$k_c = \ir/2$.
It should be noted that the fast reaction limit is reached very quickly;
in both plots, the curve closest to the fast reaction curve has a $\ir$ 
value of only $10$.
}
\end{minipage}
\ec
\end{figure}

The polarographic curves for all $\delta$
values collapse onto each other in the limit of fast reaction kinetics
(Figure~\ref{figure:fast_rxn_limits}).  Even the assumption of
symmetry in the electron-transfer reaction is not required.  When
reaction rates are much larger than the current, the two reaction-rate
terms in the Butler-Volmer equations, Eqs.~(\ref{eq:bveqc}) and
(\ref{eq:bveqa}), must balance each other at leading order:
\bea
k_c (1-\i) e^{-\zeta_o + \alpha_c(\zeta_o + \phi_o)} - \ir
e^{-\alpha_a(\zeta_o + \phi_o)}  & \approx & 0   \label{eq:bv_c_fast_rxn} \\
- k_c (1+\i) e^{-\zeta_1 + \alpha_c(\zeta_1 + \phi_1 - v)} + \ir
e^{-\alpha_a(\zeta_1 + \phi_1 - v)} & \approx & 0. 
\label{eq:bv_a_fast_rxn} 
\eea
Since $\alpha_c + \alpha_a = 1$ for theoretical models of 
single electron-transfer 
reactions~\cite{newman_book,brett_book,chidsey_article}, 
we can solve explicitly for $\phi_o$ and $v-\phi_1$ to 
obtain
\bea
  \phi_o = \ln \left ( \frac{\ir}{k_c (1-\i)} \right ) \ \ \ , \ \ \ 
  v-\phi_1 = \ln \left ( \frac{k_c (1+\i)}{\ir} \right ).
\eea
Thus, for fast reaction kinetics, the leading order cell voltage is given by 
\beq 
  v(\i) = 4 \tanh^{-1}(\i).
\eeq
Notice that this is exactly the fast reaction limit of $v(\i)$ that we find in 
both the Gouy-Chapman and Helmholtz limits.  
It is straightforward to check the validity of the assumptions made in 
Eqs.~(\ref{eq:bv_c_fast_rxn}) and (\ref{eq:bv_a_fast_rxn})
by substituting these results into the expresssions for the reaction rates 
and observing that the zeta potentials satisfy the bounds 
$\zeta_o \le 0$ and 
$\zeta_1 \le \ln \left ( \frac{k_c(1+\i)}{\ir - \i} \right )$ 
which follow from the monotonicity of $\zeta_o$ and $\zeta_1$ as functions of 
$\delta$.

\section{ Thick Double Layers, $\epsilon =O(1)$}
Up to this point, we have only examined the current-voltage
characteristics in the singular limit $\epsilon \rightarrow 0$, where
the current density cannot exceed its diffusion-limited value,
$\i=1$. The situation changes changes for any finite $\epsilon > 0$.

\subsection{What Limiting Current?}
\begin{figure}[htb]
\bc
\scalebox{0.5}{\includegraphics{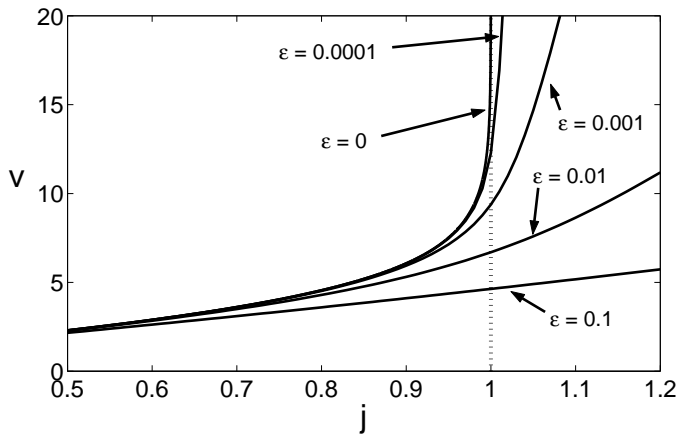}}
\begin{minipage}[h]{5in}
\caption{
\label{figure:jv_vary_eps}
Polarographic curves for $\epsilon$ values of $0$, $0.0001$, $0.001$,
$0.01$, and $0.1$ (listed in order from uppermost to lowest curves)
with the other physical parameters taken to be $\delta = 0$, $k_c=10$,
and $\ir = 10$.  Notice that for any $\epsilon > 0$, the cell has no
problem achieving current densities higher than the diffusion limited
current (dashed vertical line).  All of these curves, with the
exception of the exact $\epsilon = 0$ curve, were generated by
numerically solving Eq.~(\ref{eq:cphieq})-(\ref{eq:phieq}) subject to
the boundary conditions (\ref{eq:potbc0})-(\ref{eq:cint}) using the
method of our companion paper\cite{chu2004}.  }
\end{minipage}
\ec
\end{figure}

As is clearly evident in Figure~\ref{figure:jv_vary_eps}, the cell has
no problem breaking through the classical limiting current for
$\epsilon > 0$.  Figure~\ref{figure:jv_vary_eps} also shows that the
$\epsilon$ dependence of the polarographic curves only becomes
significant at currents approaching the diffusion-limited current;
below $\i \approx 0.5$, the curves are nearly indistinguishable.
Moreover, as $\epsilon$ increases, upper end of the polarographic
curves flatten out and shift downwards. This decrease in the cell
voltage for large $\epsilon$ values arises because the diffuse charge
layers overlap and are able to interact with each other.  More
precisely, the cell has become so small (relative to the Debye
screening length) that the electric fields from the two diffuse layers
partially cancel each other out throughout the cell resulting in a
lower total cell voltage.  It should be emphasized that this effect is
only observable because we are studying a two electrode system.
Single electrode systems (in addition to being not physically
achievable) are not capable of showing this behavior because they
always implicitly assume an infinite system size which effectively
discards any interactions from ``far away'' electrodes.

\subsection{Breakdown of the Classical Approximation} 
For a diffusion-limited cell, the classical nonlinear
asymptotic analysis just presented leads to an aesthetically appealing
theory that predicts a limiting current at $\i = 1$.  The existence of
this limiting current fits nicely with our physical intuition that the
concentration of cations in a solution must always remain nonnegative.
In reality, however, the analysis breaks down as the current
approaches (and exceeds) its limiting value.

The breakdown of the classical asymptotics is evident upon examining
the expansions for the bulk field variables as the current is
increased toward its diffusion-limited value.  Calculating a few of the
higher order terms in the bulk asymptotic expansion, we find that
\bea
  -\Eb(x) &=& \frac{2\i}{\cb^{(0)}} 
            + \frac{3}{2} \epsilon^2 \frac{(2\i)^3}{\left(\cb^{(0)}\right)^4} 
            + \frac{111}{4} \epsilon^4 \frac{(2\i)^5}{\left(\cb^{(0)}\right)^7} 
      + \frac{6045}{4} \epsilon^6 \frac{(2\i)^7}{\left(\cb^{(0)}\right)^{10}} 
      + O(\epsilon^8) 
  \label{eq:higher_order_terms_E} \\
  \cb(x) &=& \cb^{(0)}
            + \frac{1}{2} \epsilon^2 \frac{(2\i)^2}{\left(\cb^{(0)}\right)^2} 
            + \frac{3}{2} \epsilon^4 \frac{(2\i)^4}{\left(\cb^{(0)}\right)^5} 
            + \frac{231}{8} \epsilon^6 \frac{(2\i)^6}{\left(\cb^{(0)}\right)^8} 
            + O(\epsilon^8) 
  \label{eq:higher_order_terms_c} \\
  \rhob(x) &=& 0
            + \epsilon^2 \frac{(2\i)^2}{\left(\cb^{(0)}\right)^2} 
            + 6 \epsilon^4 \frac{(2\i)^4}{\left(\cb^{(0)}\right)^5} 
            + \frac{777}{4} \epsilon^6 \frac{(2\i)^6}{\left(\cb^{(0)}\right)^8} 
            + O(\epsilon^8). 
  \label{eq:higher_order_terms_rho}
\eea
Since $\cb^{(0)} \rightarrow 2x$ as $\i \rightarrow 1$, 
the higher order terms are clearly more singular than the leading order 
term at the limiting current. 
Rubinstein and Shtilman make a similar observation from a potentiostatic
perspective; they note that the asymptotic expansions are not uniform 
in the cell voltage~\cite{rubinstein1979}.
 
The inconsistency in the classical approximation was apparently first
noticed by Levich who observed that the leading-order solution in the
bulk predicts an infinite charge density when the current density
reaches $1$, which directly contradicts the assumption of bulk charge
neutrality~\cite{levich_book}.  As $\i \rightarrow 1$, the bulk charge
density is given by 
\beq \rhob =
-\epsilon^{2}\frac{d^{2}\phib}{dx^{2}} = \frac{\epsilon^2}{\left [ x +
(1-\i)/2\i \right ]^{2}} \approx \frac{\epsilon^2}{x^2}, 
\eeq 
which
diverges at the cathode.

\begin{figure}[htb]
\bc
\scalebox{0.5}{\includegraphics{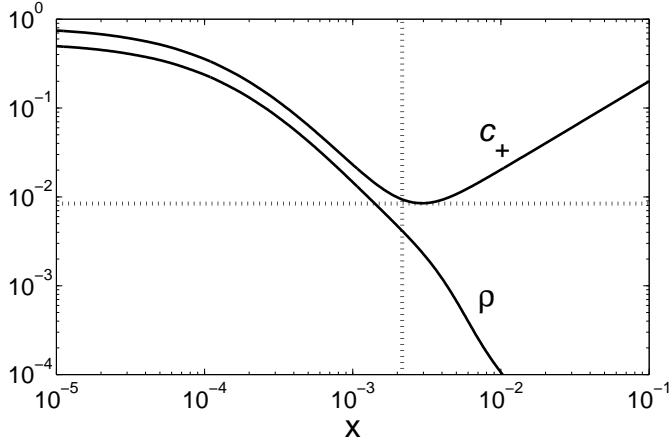}}
\begin{minipage}[h]{5in}
\caption{
\label{figure:c_rho_diffuse_layer_expansion}
Numerical solutions for the dimensionless cation concentration $c_+(x)$
and (full) charge density $2 \rho(x)$ at the diffusion-limited current 
($j=1.0$) with physical parameters $k_c = 10$, $\ir = 10$, 
$\alpha_c=\alpha_a=0.5$, $\delta=0.0$ and $\epsilon = 0.0001$.
In the cathode region, electroneutrality breaks down as the solution
becomes cation rich in order to satisfy the reaction boundary 
conditions.  Note that when $x=O(\epsilon^{2/3})$, $c_+$ and $\rho$ are
both $O(\epsilon^{2/3})$. 
For reference, the dashed vertical line shows where $x = \epsilon^{2/3}$,
and the dashed horizontal line shows where 
$y = \epsilon^{2/3} \left[ (2+2^{2/3}) + 4/(2+2^{2/3})^2 \right] 
\approx c_+(\epsilon^{2/3})$. 
}
\end{minipage}
\ec
\end{figure}

Smyrl and Newman first showed that these paradoxical results are
related to the breakdown of thermal equilibrium charge profiles near
the cathode, leading to a significant expansion of the double layer
into the bulk solution~\cite{smyrl1967}.  They argue that the
assumption of electroneutrality breaks down when $\rhob \approx \cb$.
Since $\cb$ is proportional to $x$ at the limiting current, the bulk
approximation fails to be valid for $x$ smaller than
$O(\epsilon^{2/3})$, which leads to a boundary layer that is thicker
than the usual Debye length.  From an alternative perspective, the
problems begin when $\rhob(0)/\cb(0) \approx 1$ (see
Figure~\ref{figure:c_rho_diffuse_layer_expansion}).  
Using this
criterion, we find that the classical asymptotic theory is only
appropriate when $c_o \gg (2 \i \epsilon)^{2/3}$ or, equivalently, $\i
\ll 1 - (2 \epsilon)^{2/3}$.  Since the cell voltage is approximately
$4 \tanh^{-1}(\i)$ in many situations, this regime also corresponds to
$v = O\left ( | \ln \epsilon | \right )$. This shows that in thin
films, where $\epsilon$ is not so small, it is easy to
exceed the classical
limiting current and achieve rather different charge profiles~\cite{chu2004}.

\section{Conclusion}  
In summary, we have revisited the classical PNP equations, analyzing
for the first time the effect of physically realistic boundary
conditions for thin-film galvanic cells and other
micro-electrochemical systems. In particular, we focus on the effect
of Stern-layer capacitance and Faradaic reactions with Butler-Volmer
kinetics. Such boundary conditions contain new physics, such as the
possibility of a reaction-limited current due to the slow injection of
ions at the anode.  We also find that the Stern layer generally allows
the cell to exceed limiting currents by carrying diverging portions of
the cell voltage, which would otherwise end up in the diffuse part of
the double layer. We have provided analytical formulae for
current-voltage relations that should provide useful in characterizing
the differential resistance of thin films, such as those used in
on-chip micro-batteries. Here, we have focused on the classical
nonlinear regime in which thin double layers remain in thermal
equilibrium; the more exotic, non-equilibrium regime, which arises at
and above the classical limiting current, is analyzed in the companion
paper~\cite{chu2004}.

\section{Acknowledgments} 
This work was supported in part by the MRSEC Program of the National
      Science Foundation under award number DMR 02-13282 and in part
      by the Department of Energy through the Computational Science
      Graduate Fellowship (CSGF) Program provided under grant number
      DE-FG02-97ER25308. The authors thank F. Argoul, J. J. Chae,
      H. A. Stone, and W. Y. Tam, for many helpful discussions.

\appendix

\section{Positivity of Ion Concentrations \label{appendix:A}} 
The positivity of the ion concentrations follows directly from the 
mathematical formulation of the problem.  For the anion concentration,
equations (\ref{eq:c-eq}) can be integrated exactly using the integrating 
factor $e^{\phi}$ to yield
\beq
c_-(x) = A e^{\phi(x)},
\eeq
which implies that the sign of $c_-(x)$ is the same across the entire domain.
Since the integral constraint (\ref{eq:cint}) requires that $c_-(x)$ is
positive somewhere in the domain, $c_-(x)$ must be positive {\it everywhere}
in the domain.

For the cation concentration, we make use of the reaction boundary conditions.
Integrating equation (\ref{eq:c+eq}) with the integrating factor $e^{\phi}$, 
we obtain
\beq
c_+(x) = 
  c_+(0) e^{\phi(0)-\phi(x)} + 4 \i e^{-\phi(x)}\int_{0}^{x} e^{\phi(y)} dy .
\eeq
Clearly, the integral term is postive because $e^\phi$ is positive everywhere.
Moreover, the reaction boundary condition (\ref{eq:potbvbc0}) implies that 
$c_+(0) > 0$ because both $\ir$ and $k_c$ are positive.  Thus, we find that
the cation concentration is strictly positive.

\end{document}